\documentclass[aps,prc,twocolumn,longbibliography,reprint,superscriptaddress,amsmath,nofootinbib]{revtex4-1}

\usepackage{verbatim}
\usepackage{amsmath,amssymb,amsfonts}
\usepackage{graphicx}
\usepackage[dvipsnames]{xcolor}
\usepackage[colorlinks=true,linkcolor=black,urlcolor=blue,citecolor=blue]{hyperref}
\usepackage[normalem]{ulem}

\graphicspath{{Figs/}}

\usepackage{bm} 

\newcommand{\feq}{f_\mathrm{eq}}
\newcommand{\fME}{f_\mathrm{ME}}
\newcommand{\mVAH}{\texttt{mVAH}}

\begin{document}

\title{Fluid dynamics from the Boltzmann equation using a maximum entropy distribution}

\author{Chandrodoy Chattopadhyay}
\affiliation{Department of Physics, North Carolina State University, Raleigh, NC 27695, USA}
\author{Ulrich Heinz}
\affiliation{Department of Physics, The Ohio State University, Columbus, OH 43210, USA}
\author{Thomas Sch\"afer}
\affiliation{Department of Physics, North Carolina State University, Raleigh, NC 27695, USA}

\date{\normalsize \today}

\begin{abstract}
\noindent
Using the recently developed ‘Maximum Entropy’ (or ‘least biased’) distribution function to truncate the moment hierarchy arising from kinetic theory, we formulate a far-from-equilibrium macroscopic theory that provides the possibility of describing both free-streaming and hydrodynamic regimes of heavy-ion collisions within a single framework. Unlike traditional hydrodynamic theories that include viscous corrections to finite order, the present formulation incorporates contributions to all orders in shear and bulk inverse Reynolds numbers, allowing it to handle large dissipative fluxes. By considering flow profiles relevant for heavy-ion collisions (Bjorken and Gubser flows), we demonstrate that the present approach provides excellent agreement with underlying kinetic theory throughout the fluid’s evolution and, especially, in far-off-equilibrium regimes where traditional hydrodynamics breaks down.
\end{abstract}

\maketitle

\section{Introduction}
\label{sec1}
%

Obtaining equations of hydrodynamics, a macroscopic theory that governs the space-time evolution of conserved densities of a system, from a microscopic description of the same system requires `coarse-graining'. For a system composed of weakly-interacting particles described kinetically in terms of a single-particle phase-space distribution $f(x,p)$, the coarse-graining can be achieved by integrating out the momentum information encoded in the distribution and focusing the attention on its lowest momentum moments. In kinetic theory, the time-evolution of the particle distribution is described by the Boltzmann equation which can be re-cast as an infinite hierarchy of equations for the coarse-grained `moments’ of the distribution. The low-order moments correspond to conserved densities, their fluxes, as well as non-equilibrium components of the conserved currents, such as shear and bulk viscous stresses and charge diffusion currents. In the full microscopic description, the evolution of these non-equilibrium fluxes couples to higher-order `non-hydrodynamic’ moments of the distribution. To obtain a hydrodynamic theory one therefore requires a truncation scheme, i.e. a procedure to close the system of equations by expressing the non-hydrodynamic moments in terms of hydrodynamic ones. This step amounts to re-constructing an approximate distribution function using only information contained in a handful of its low-order moments. While this procedure is crucial in determining the range of applicability of the ensuing hydrodynamic equations, it is inherently ambiguous and mathematically ill-defined. The main objective of this paper is to use a truncation distribution that, on the one hand, originates from a well-motivated information-theoretical principle, and on the other allows for the formulation of a hydrodynamic theory that can also work when the system is not close to local thermal equilibrium. 

Previous work on obtaining relativistic hydrodynamics from kinetic theory explored several different classes of truncation distributions. Well-known examples are the Grad 14-moment approximation \cite{Grad, Muller:1967zza, Israel:1976tn, Israel:1979wp}, the Chapman-Enskog expansion \cite{Chapman:1970}, and the more recently introduced Romatschke-Strickland anisotropic distribution \cite{Romatschke:2003ms, Tinti:2015xwa}. All these truncation distributions invoke extra, and sometimes ad-hoc, approximations that are based on specific assumptions about the microscopic kinetic dynamics and thereby introduce additional information on top of that contained in the hydrodynamic degrees of freedom. For example, in Grad’s approach, the distribution function is expanded around a local thermal form (J\"uttner distribution) in powers of particle momenta. This was extended to relativistic kinetic theory by Israel and Stewart \cite{Israel:1976tn,Israel:1979wp} and, more recently, by Denicol, Niemi, Molnar, and Rischke \cite{Denicol:2012cn} into a general framework for second-order transient relativistic fluid dynamics. A slightly different approximation for the truncating distribution function where the expansion is in terms of the fluid’s velocity gradients (Chapman-Enskog series \cite{Chapman:1970}), was implemented by Jaiswal \cite{Jaiswal:2013npa, Jaiswal:2013vta}. Both of these approaches assume proximity of the fluid to local equilibrium and lead to inconsistencies when the system is far from equilibrium. A typical manifestation of the latter is that the distribution function turns negative (unphysical) at large particle momenta. Some of these problems can be by-passed by using the Romatschke-Strickland ansatz as a truncation distribution. Doing so gives rise to `anisotropic hydrodynamics’ and its extensions \cite{Florkowski:2010cf, Martinez:2010sc, Martinez:2012tu, Bazow:2013ifa, Florkowski:2014bba, Molnar:2016vvu, Molnar:2016gwq, Alqahtani:2017mhy, McNelis:2018jho, Nopoush:2019vqc}. Although the resulting framework does not require the fluid to be close to local equilibrium, this approach assumes a rather particular flow pattern, namely, the early stage dynamics of quark-gluon plasma formed in relativistic heavy-ion collisions.\footnote{%
    For a review of applications of hydrodynamics to heavy-ion collisions, the reader may refer to \cite{Heinz:2013th}.
}
In such collisions the fluid initially undergoes strongly anisotropic expansion mostly along the beam direction, leading to substantial momentum space anisotropy which the Romatschke-Strickland ansatz \cite{Romatschke:2003ms} captures efficiently. However, this begs the question how to generalize this ansatz in a principled approach \cite{Tinti:2015xwa} to far-off-equilibrium fluids expanding with more general, fully three-dimensional flow profiles.

Recently, Everett {\it et al.} \cite{Everett:2021ulz} proposed the Maximum Entropy principle \cite{Jaynes:1957zza, Jaynes:1957zz} as the guiding idea for reconstructing a single-particle distribution function uniquely using only macroscopic information encoded in the hydrodynamic variables of a relativistic system. The resulting distribution, known as the `maximum-entropy’ or `least-biased’ distribution, maximizes the Shannon (information) entropy of the system subject to all, but nothing else than, the information contained in the hydrodynamic degrees of freedom. The work in  \cite{Everett:2021ulz} was motivated by the problem of `particlizing' the fluid in a heavy-ion collision at the end of its evolution. This procedure provides the dsitribution of particles emitted from the collision fireball, including their species identity, space-time positions, and four-momenta.\footnote{%
    It was recently generalized by Pradeep and Stephanov \cite{Pradeep:2022eil} to uniquely reconstruct multiparticle distribution functions that correctly encode also the thermal and critical fluctuations inherited from the hydrodynamic evolution of the fluid.}
In the present work we generalize this approach by turning it into a dynamical framework for general far-off-equilibrium hydrodynamic evolution in 3+1 dimensions, which we call {\it ME hydrodynamics} (or {\it ME hydro} in short). 

In essence, ME hydro implements the following algorithm: The task is to evolve the energy-momentum tensor $T^{\mu\nu}$ and the conserved baryon, strangeness and electric currents $j^\mu_{B,S,Q}$ in space and time. We use hydrodynamic evolution equations obtained from momentum moments of the relativistic Boltzmann equation. There are two types of equations: the conservation laws for energy, momentum and the conserved charges, and a set of relaxation-type equations for the dissipative flows (i.e. the bulk and shear viscous stresses and that diffusion currents). The latter couple to higher, non-hydrodynamic moments of the distribution function and therefore require truncation. The form of these equations is universal \cite{Heinz:2015gka} -- detailed information about the microscopic physics in the medium is encoded in its equation of state and transport coefficients. 

When advancing the system by one time step, most terms in the evolution equations can be obtained from the components of $T^{\mu\nu}$ and $j^\mu_{B,S,Q}$ from the previous time step. The terms which can not, i.e. the couplings to higher, non-hydrodynamic modes (but only these!), are evaluated with the following simple model: Assume that the fluid is made up of a (single- or multi-component, as desired) gas of massive particles with Boltzmann, Bose-Einstein and/or Fermi-Dirac statistics (again as desired), with mass(es) corresponding to typical expectations for the presumed microscopic constituents. Match an equilibrium distribution $\feq$ and a Maximum Entropy distribution $\fME$ for the gas particles to $T^{\mu\nu}$ and $j^\mu_{B,S,Q}$ from the previous time step, and use $\delta f = \fME-\feq$ to evaluate the non-hydrodynamic moments. Advance by one step, and repeat.

It should be noted that the Maximum Entropy distribution $\fME$ introduced in this algorithm for the purpose of truncating the moment hierarchy {\it does not} solve the Boltzmann equation, nor is its deviation from equilibrium $\delta f = \fME-\feq$ assumed to be small. It is just the `most likely' $\delta f$ (in an information theoretical sense) compatible with the hydrodynamic state at each time step and at every point of the spatial grid. The described procedure is unambiguous and well-motivated even for large dissipative flows (i.e. large inverse Reynolds numbers Re$^{-1}$).

In addition to the fact that the Maximum Entropy distribution $\fME$ stems from a deep connection between information theory and statistical mechanics, it has several other appealing features. For example, it is positive definite over the entire range of particle momenta,\footnote{%
    This allows it to be interpreted as a probability density of particles in phase-space.
}
it can generate a wide range of non-equilibrium stresses allowed by kinetic theory, and it does not require the system to be close to local thermal equilibrium. Accordingly, we propose to truncate the infinite hierarchy of moment equations for the Boltzmann equation with $\fME$, in order to obtain hydrodynamic equations that can be used even far away from local equilibrium.

For non-relativistic fluids this method was introduced in the 1990's by Levermore \cite{Levermore} who showed that using a Maximum Entropy distribution for moment closure leads to a system of equations that is hyperbolic, i.e. its initial value problem is well-posed, and that the second law of thermodynamics is always satisfied. Murchikova {\it et al.} studied and compared it to other truncation schemes in Ref.~\cite{Murchikova:2017zsy} for the relativistic Boltzmann equation for neutrino transport in astrophysics. For conformally invariant relativistic fluids a related approach called `dissipative type theory' (DTT) was advocated  by Cantarutti and Calzetta \cite{Calzetta:2010au, Peralta-Ramos:2012tgz, Calzetta:2019dfr}. However, in contrast to ME hydrodynamics, DTT is not based on the Maximum Entropy principle, but on the principle of maximizing {\it the rate of entropy production}. Since that rate is controlled by the collision term in the Boltzmann equation, maximizing it requires additional information about the microscopic kinetic processes which ME hydrodynamics does not require.\footnote{%
    As will be discussed in Section~\ref{sec_mehydro_intro}, in deriving DTT the authors of \cite{Calzetta:2019dfr} make several additional approximations, one of which is the assumption of small deviations from local equilibrium. As a result of these approximations their distribution $f_{DTT}$ agrees exactly with $\fME$ but, according to the approximations made, their form for $f_{DTT}$ should not be used in far-from-equilibrium situations.
    }

This paper is organized as follows. In Sec. \ref{sec1_old} we review the relativistic Boltzmann equation with a relaxation-type collisional kernel and its re-formulation in terms of moment equations. We then briefly describe in Secs. \ref{sec_Grad_intro}-\ref{sec_mehydro_intro} some of the standard truncation schemes, culminating in the Maximum Entropy truncation procedure. In Secs.~\ref{sec_Bjorken_main} and \ref{sec3} we test the performance of ME hydrodynamics in two highly symmetric situations, Bjorken and Gubser flow, for which the underlying kinetic theory can be solved exactly, allowing for a quantitatively precise comparison between the microscopic and macroscopic descriptions. While the symmetries underlying Gubser flow include conformal symmetry, this symmetry can be broken in Bjorken flow for which we study both massless and massive constituents. Both of these flow profiles include regimes where the system is far-from-equilibrium. Thus they provide stringent test beds for ME hydrodynamics far from local equilibrium. For both flow profiles ME hydrodynamics passes the test with flying colors. Conclusions are offered in Sec.~\ref{sec_conclusions}. Several appendices add technical details and provide further clarifications. 

\section{The Boltzmann equation in relaxation time approximation}
\label{sec1_old}

We consider a weakly coupled system of particles whose dynamics is described statistically by relativistic kinetic theory. The statistical description relies on a single-particle distribution function $f(x,p)$ which gives the mean density of particles in phase-space. The evolution of the distribution function is governed by the relativistic Boltzmann equation
\begin{align}\label{BE}
    p^\mu \partial_\mu f(x,p) = {\cal C}[f], 
\end{align}
where $p^\mu$ is the particle four-momentum with $p^\mu p_\mu = m^2$, $m$ being the particle mass. The collisional kernel ${\cal C}[f]$ models the interactions between particles and typically includes the effects of $2 \leftrightarrow 2$ scatterings. In this work, we choose a simplistic collisional kernel given by the relaxation-time approximation (RTA) \cite{anderson1974relativistic} where the complicated effects of interactions are assumed to drive the system toward \textit{local} equilibrium. To specify local equilibrium, this description introduces macroscopic variables like flow velocity $u^\mu(x)$, temperature $T(x)$, and chemical potential $\mu(x)$ such that the collisonal kernel is approximated by \cite{anderson1974relativistic} 
\begin{align}\label{RTA_Cf}
    {\cal C}[f] \approx - \frac{u \cdot p}{\tau_R} \, \left( f - \feq \right).
\end{align}
In the above, $\tau_R(x)$ is the relaxation-time whose functional dependence on temperature and chemical potential has to be parametrized. The form of the equilibrium phase-space function is given by the J\"uttner distribution
\begin{align}
\feq(x,p) = \Bigl\{\exp\bigl[ (u(x) \cdot p)/T(x) - \alpha(x) \bigr] - \epsilon \Bigr\}^{-1},
\end{align}
where $u^\mu(x)$, with $u^\mu u_\mu=1$, is the four-velocity of the local fluid rest frame, $\alpha(x) \equiv \mu(x)/T(x)$ is the reduced chemical potential, and $\epsilon = -1, 0, 1$ distinguishes between Fermi-Dirac, Maxwell-Boltzmann, or Bose-Einstein statistics. In the following, we assume that the particles satisfy Boltzmann statistics ($\epsilon = 0$) and their number is not conserved ($\alpha(x) = 0$). This reduces the number of macroscopic variables to 4, namely, $u^\mu(x)$ and $T(x)$. These variables are determined by demanding that the RTA collisional kernel (\ref{RTA_Cf}) satisfies energy-momentum conservation. The energy-momentum tensor is the second-moment of $f(x,p)$,
\begin{align}\label{Tmunu_kin}
    T^{\mu\nu} \equiv \langle p^\mu \, p^\nu \rangle, 
\end{align}
where we use the notation $\langle \cdots \rangle \equiv \int dP \, ( \cdots ) \, f(x,p)$, with $dP \equiv d^3p/[(2\pi^3) E_p]$ being the Lorentz invariant phase-space measure and $E_p = \sqrt{p^2 + m^2}$ the particle energy. We also define $\langle \cdots \rangle_\mathrm{eq} \equiv \int dP \, (\cdots) \, \feq$ and $\langle \cdots \rangle_\delta \equiv \int dP \, (\cdots) \, \delta f $ where $\delta f \equiv f - \feq$ is the deviation from local equilibrium. Demanding $\partial_\mu T^{\mu\nu} = 0$ and using the Boltzmann equation with ${\cal C}[f]$ given by Eq.~(\ref{RTA_Cf}) for a momentum-independent $\tau_R$ yields\footnote{%
    If the relaxation-time depends on particle momentum, the RTA collisional kernel has to be generalized to be compatible with conservation laws \cite{Teaney:2013gca, Rocha:2021zcw, Rocha:2021lze, Dash:2021ibx}.
} 
\begin{align}\label{Landau_matching1}
    T^\mu_\nu \, u^\nu = e_\mathrm{eq} \, u^\mu
\end{align}
where for our system the equilibrium energy density $e_\mathrm{eq} = \bigl\langle \left( u \cdot p \right)^2 \bigr\rangle_\mathrm{eq}$ is given by the equation of state (EoS)
\begin{align}\label{e_eq}
    e_\mathrm{eq}(T,m) &= 
     \frac{3T^4 z^2}{2\pi^2} \, \left( K_2(z) + \frac{z}{3} K_1(z) \right),
\end{align}
with $z \equiv m/T$ and $K_n$ being the modified Bessel functions of second kind of order $n$. Eq.~(\ref{Landau_matching1}) is the Landau matching condition, stating that $u^\nu$ and $e_\mathrm{eq}$ are the time-like eigenvector and associated eigenvalue of $T^{\mu}_\nu$. The EoS (\ref{e_eq})  defines the local temperature $T(x)$ in terms of the energy density $e(x)$ in the fluid's rest frame, i.e., $e(x) = e_{\mathrm{eq}}(T(x), m)$. The matching condition (\ref{Landau_matching1}) implies that the energy-momentum tensor of the system can be decomposed as
\begin{align}
\label{Tmunu}
    T^{\mu\nu} 
    = e_\mathrm{eq} \, u^\mu \, u^\nu - \left( P + \Pi \right) \, \Delta^{\mu\nu} + \pi^{\mu\nu},
\end{align}
where $\Delta^{\mu\nu} \equiv g^{\mu\nu} - u^\mu u^\nu$ projects any tensor orthogonal to $u^\mu$. The coefficient of $\Delta^{\mu\nu}$ is the total isotropic pressure which is split into an equilibrium part, $P(T,m)$, and a bulk viscous part $\Pi$. The symmetric, traceless, and orthogonal (to $u^\mu$) part of $T^{\mu\nu}$ is the shear stress tensor $\pi^{\mu\nu}$. In local equilibrium both $\Pi$ and $\pi^{\mu\nu}$ vanish, i.e. they arise solely from deviations from local equilibrium:
\begin{align}
    \Pi &\equiv - \frac{1}{3} \, \langle \Delta_{\mu\nu} \, p^\mu \, p^\nu \rangle_\delta, \nonumber \\
    \pi^{\mu\nu} &\equiv \langle p^{\langle \mu} \, p^{\nu \rangle} \, \rangle_\delta.
    \label{bulk_shear_definitions}
\end{align}
In the last definition we used the notation $A^{\langle \mu \nu \rangle} \equiv \Delta^{\mu\nu}_{\alpha\beta} \, A^{\alpha\beta}$, with the double-symmetric, traceless and orthogonal projector 
\begin{align}\label{projector}
    \Delta^{\mu\nu}_{\alpha\beta} \equiv  \left( \Delta^\mu_\alpha \, \Delta^\nu_\beta + \Delta^\mu_\beta \, \Delta^\nu_\alpha \right)/2   - \frac{1}{3} \, \Delta^{\mu\nu} \, \Delta_{\alpha\beta}.
\end{align}
Using Eq.~(\ref{Tmunu_kin}) it is straightforward to show that the Landau matching condition  (\ref{Landau_matching1}) implies
\begin{align}\label{Landau_matching2}
    \bigl\langle \left( u \cdot p \right)^2 \bigr\rangle_\delta = 0, \quad 
    \bigl\langle p^{\langle \mu \rangle} \, \left( u \cdot p \right) \bigr\rangle_\delta = 0.  
\end{align}

One can either solve the transport equation  (\ref{BE}) to obtain $f(x,p)$ and then use it to calculate the shear and bulk viscous pressures from Eq.~(\ref{bulk_shear_definitions}), or re-cast the Boltzmann equation into an infinite system of coupled evolution equations for moments of $f(x,p)$ whose solution yields the bulk and shear stresses. Exact evolution equations for the dissipative fluxes $\Pi$ and $\pi^{\mu\nu}$ were obtained by Denicol, Niemi, Molnar, and Rischke \cite{Denicol:2012cn}. For a single-component Boltzmann gas of particles with mass $m$, without conserved currents (i.e. with zero chemical potential), the bulk and shear evolution equations are \cite{Denicol:2014vaa, Denicol:2012cn}:
\begin{align}
    &\dot{\Pi} + \frac{\Pi}{\tau_R} = - \beta_\Pi \, \theta - \left( 1 - c_s^2 \right) \, \Pi \, \theta + \frac{m^4}{9} \, \rho_{-2} \, \theta \nonumber \\
    & + \left( \frac{1}{3} - c_s^2 \right) \, \pi^{\mu\nu} \, \sigma_{\mu\nu} + \frac{m^2}{3} \, \rho^{\mu\nu}_{-2} \, \sigma_{\mu\nu}, \label{bulk_Denicol} \\ 
    &\dot{\pi}^{\langle \mu\nu \rangle} + \frac{\pi^{\mu\nu}}{\tau_R} = 2 \beta_\pi \sigma^{\mu\nu} - \frac{10}{7} \, \pi^{\mu\langle \lambda} \, \sigma^{\nu\rangle}_\lambda 
    + 2 \, \pi^{\lambda \langle \mu} \omega^{\nu \rangle}_{\lambda} \nonumber  \\
    & - \frac{4}{3} \, \pi^{\mu\nu} \, \theta + \frac{4 m^2}{7} \rho_{-2}^{\mu \langle \lambda} \sigma^{\nu \rangle}_{\lambda} - \rho_{-2}^{\mu\nu\lambda\rho} \sigma_{\lambda\rho} - \Delta^{\mu\nu}_{\alpha\beta} \nabla_{\lambda} \rho^{\alpha\beta\lambda}_{-1} \nonumber\\
    & + \frac{2}{5} \nabla^{\langle \mu} \left( \rho^{\nu\rangle}_1 - m^2 \rho^{\nu \rangle}_{-1} \right) - 2 \rho_{1}^{\langle \mu} \dot{u}^{\nu \rangle} - \frac{m^2}{3} \rho_{-2}^{\mu\nu} \theta \nonumber \\ 
    &  + \left( \frac{6}{5} \, \Pi - \frac{2}{15} \, m^2 \,  \rho_{-2}\right) \sigma^{\mu\nu}. \label{shear_Denicol}
\end{align}
Here $\dot{A} \equiv u^\mu \partial_\mu A$ denotes the time derivative and $\nabla^\mu \equiv \Delta^{\mu\alpha}\partial_\alpha$ the spatial gradient in the local fluid rest frame, $\sigma^{\mu\nu} \equiv \nabla^{\langle \mu} u^{\nu\rangle}$ is the velocity shear tensor, and $\theta \equiv \partial_\mu u^\mu$ is the scalar expansion rate. The speed of sound $c_s^2$ is a so-called zeroth-order transport coefficient,
\begin{align}
    c_s^2 \equiv dP/de =  \frac{e + P}{3e + \left( 3 + z^2 \right) P},
\end{align}
with $z \equiv m/T$. $\beta_\Pi$ and $\beta_\pi$ are defined as \cite{Denicol:2014vaa}
\begin{align}
    \beta_\Pi &= \left( \frac{1}{3} - c_s^2 \right) \left( e + P \right) - \frac{2}{9} \left( e - 3 P \right) - \frac{m^4}{9} \, I_{-2,0}, \\
    \beta_\pi &= \frac{4 P}{5} + \frac{1}{15} \left( e - 3 P \right) - \frac{m^4}{15} \, I_{-2, 0},
\end{align}
where
\begin{align}
    \!\!\!\!
    I_{n,q} \equiv \frac{1}{(2q{+}1)!!} \, \int dP \, \left( u \cdot p \right)^{n-2q} \, \left( \Delta_{\mu\nu} \, p^\mu \, p^\nu  \right)^q \, \feq .
\end{align}
They are related to the first-order bulk and shear viscosity coefficients $\zeta \equiv \tau_R \, \beta_\Pi$ and $\eta = \tau_R \, \beta_\pi$.

Eqs.~(\ref{bulk_Denicol},\ref{shear_Denicol}) are not closed, because of couplings to the $\rho$-moments which stem from the non-equilibrium part of the distribution function,
\begin{align}
    \rho^{\mu_1\cdots \mu_l}_n \equiv \langle \left( u \cdot p \right)^n \, p^{\langle \mu_1} \cdots \, p^{\mu_l \rangle} \rangle_\delta ,
\end{align}
where
\begin{align}
    p^{\langle \mu_1} \cdots p^{\mu_l \rangle} \equiv \Delta^{\mu_1 \cdots \mu_l}_{\nu_1 \cdots \nu_l} \, p^{\nu_1} \cdots p^{\nu_l}.
\end{align}
The spatial (in the local rest frame) projectors $\Delta^{\mu_1 \cdots \mu_l}_{\nu_1 \cdots \nu_l}$ are defined in Appendix F of Denicol {\it et al.} \cite{Denicol:2012cn}; for $l=2$ they reduce to the expression (\ref{projector}).

In many situations it is of interest to consider the long-wavelength, slow dynamics of the underlying microscopic evolution. In general, the long-wavelength description of any macroscopic medium is governed by hydrodynamics, which is based on the conservation of energy-momentum tensor and conserved charged currents inherent to the system. In the traditional hydrodynamic approach, the conserved tensors are solely expressed in terms of a few thermodynamic variables (temperature, chemical potentials etc.), a flow velocity, and their spatial gradients \cite{Eckart:1940zz, Landau:1987}. The constitutive relations expressing the conserved currents in terms of thermodynamic/macroscopic fields typically involve a systematic expansion in powers of velocity gradients \cite{Muronga:2001zk, Baier:2007ix, Bhattacharyya:2007vjd}. Accordingly, the traditional approach breaks down whenever these gradients, characterising the macroscopic length-scale of the problem, become large in comparison to the microscopic scale, i.e, the mean-free path of the system. Moreover, an expansion solely in powers of velocity gradients results in a framework that is acausal and numerically unstable \cite{Hiscock:1983zz, Hiscock:1985zz}. 

A dynamical description of the conserved currents that is known to provide much better agreement with the underlying microscopic theory is provided by `transient' or `relaxation-type' hydrodynamics where the out-of-equilibrium components of $T^{\mu\nu}$ (namely $\Pi$ and $\pi^{\mu\nu}$) are treated as independent dynamical degrees of freedom \cite{Muller:1967zza, Israel:1976tn, Israel:1979wp, Denicol:2012cn, Jaiswal:2013npa, Jaiswal:2013vta, Florkowski:2015lra, Grozdanov:2015kqa}. The corresponding evolution equations are obtained by truncating Eqs. (\ref{bulk_Denicol},\ref{shear_Denicol}), i.e., by expressing the `non-hydrodynamic' $\rho-$ moments \footnote{%
    They are called `non-hydrodynamic' as they do not appear in the conserved currents that hydrodynamics usually describes.
    } 
in terms of bulk and shear stresses. There are various truncation schemes, all of which approximate the distribution function $f(x,p)$ using the macroscopic quantities that appear on the r.h.s. of Eq.~(\ref{Tmunu}). In the following, we briefly mention some of the commonly used truncation procedures to obtain relaxation-type hydrodynamics.

\subsection{The 14-moment approximation}
\label{sec_Grad_intro}

In this approach, due to Grad \cite{Grad}, M\"uller \cite{Muller:1967zza}, and Israel and Stewart \cite{Israel:1976tn, Israel:1979wp}, the distribution function is expanded in powers of particle 4-momenta around the J\"uttner distribution such that $\delta f$ has 14 undetermined coefficients $(a, b_\mu, c_{\mu\nu})$\footnote{%
    Without loss of generality, $c_{\mu\nu}$ can be taken to be symmetric and traceless with its trace absorbed in the definition of `a'; this sets its number of independent coefficients to 9.    
}:
\begin{align}
    \frac{\delta f(x,p)}{\feq} & \approx a + b_\mu \, p^\mu + c_{\mu\nu}  \, p^\mu \, p^\nu, \nonumber \\
    & \equiv {\cal A} + {\cal B} \, (u\cdot p) + p^{\langle \mu \rangle} \left( b_{\langle \mu \rangle} + 2 \, (u \cdot p) \, c_{\langle\mu\rangle} \right) \nonumber \\
    & + {\cal C} \, (u\cdot p)^2 + c_{\langle \mu\nu \rangle} \, p^{\langle \mu} \, p^{\nu \rangle},  
\end{align}
where, in the second line, we defined ${\cal B} \equiv u_\mu \, b^\mu$, $c_{\langle \mu \rangle} \equiv \Delta_{\mu}^{\alpha} \, c_{\alpha\beta} \, u^\beta$, $c \equiv c_{\mu\nu} \, u^\mu \, u^\nu$, ${\cal A} \equiv a - m^2 \, c/3$, and ${\cal C} \equiv 4 c/3$. In the absence of a conserved charge current, the coefficients $b_\mu$ are set to zero and the 14-moment approximation essentially reduces to a 10-moment approximation:
\begin{align}\label{Grad_deltaf}
  \frac{\delta f(x,p)}{\feq} \approx & {\cal A} +  2 \, c_{\langle \mu \rangle} \, \left( u \cdot p \right) \, p^{\langle\mu\rangle} + {\cal C} \, (u\cdot p)^2 \nonumber \\
  & + c_{\langle \mu\nu \rangle} \, p^{\langle \mu} \, p^{\nu \rangle}.
\end{align}
The ten coefficients are determined by (i) imposing the Landau matching conditions (Eq. (\ref{Landau_matching2})) and (ii) the criteria that $\delta f$ given by Eq. (\ref{Grad_deltaf}) yields the bulk and shear stresses as per their definitions in Eqs. (\ref{bulk_shear_definitions}). Note that the Landau matching condition, $\langle p^{\langle \mu \rangle} \, (u \cdot p) \rangle_\delta = 0$ implies that $c_{\langle \mu \rangle}$ in Eq.~(\ref{Grad_deltaf}) must be zero.

\subsection{The Chapman-Enskog approximation} \label{sec_CE_intro}

In this framework, the RTA Boltzmann equation is approximately solved assuming that the Knudsen number (a ratio of microscopic scattering rate and macroscopic length scale) is small \cite{Chapman:1970}. One re-writes Eq. (\ref{RTA}) as,
\begin{align}\label{RTA_recast}
    f = \feq - \frac{ \varepsilon \tau_R}{u \cdot p} \, p^\mu \partial_\mu f,
\end{align}
where $\varepsilon$ is a power-counting parameter that signifies the smallness of the second-term. Assuming a perturbative solution of the form $ f = \sum_{i=0}^{\infty} \, f_i \, \varepsilon^i$,  and plugging it in both sides of Eq. (\ref{RTA_recast}) one obtains,
\begin{align}
    f_0 &= \feq, \,\, f_1 = - \frac{\tau_R}{u\cdot p} \, p^\mu \partial_\mu \feq, \nonumber \\
    f_2 &= \frac{\tau_R}{u\cdot p} \, p^\mu \partial_\mu \, \left( \frac{\tau_R}{u\cdot p} p^\nu \partial_\nu \feq \right),
\end{align}
and so on. Writing $f = \feq + \delta f$ and keeping terms up to first-order in velocity gradients ($\delta f \approx f_1$) the out-of-equilibrium part of the distribution function is \cite{Jaiswal:2014isa},
\begin{align}\label{deltaf_gradient}
    \delta f = \frac{\beta \, \tau_R}{u \cdot p} \, & \Bigg[ \frac{1}{3} \left\{ m^2 - \left(1 - 3 c_s^2 \right) \left( u \cdot p \right)^2 \right\} \theta \nonumber \\
    & + p^\mu \, p^\nu \, \sigma_{\mu\nu} \Bigg] \, \feq. 
\end{align}
This yields the first-order expressions of the bulk viscous pressures and shear stress tensor,
\begin{align}\label{bulk_shear_NS}
    \Pi = - \tau_R \, \beta_\Pi \, \theta, \,\, \pi^{\mu\nu} = 2 \tau_R \, \beta_\pi \, \sigma^{\mu\nu}.
\end{align}
It is customary to express $\delta f$ in terms of the shear and bulk viscous pressures instead of the scalar expansion rate and velocity stress tensor appearing in Eq. (\ref{deltaf_gradient}). For this, one uses the first-order expressions (Eq. (\ref{bulk_shear_NS})) to get \cite{Jaiswal:2014isa},
\begin{align}\label{deltaf_CE}
    \frac{\delta f}{\feq} & = - \frac{\beta}{3 (u \cdot p) \beta_\Pi} \,  \left[ m^2 - \left( 1 - 3 c_s^2 \right) \, \left( u \cdot p \right)^2  \right] \, \Pi \nonumber \\ 
    & + \frac{\beta}{2 (u \cdot p) \beta_\pi} \, p^\mu \, p^\nu \, \pi_{\mu\nu}.
\end{align}
The above form of off-equilibrium correction will be referred to as Chapman-Enskog $\delta f$ and the hydrodynamics derived from it as second-order CE-hydro. 

\subsection{The anisotropic-hydrodynamic approximation}
\label{sec_ahydro_intro}

Both the Grad's approximation as well as Chapman-Enskog $\delta f$ have certain shortcomings. The Grad's 14-moment assumption is ad-hoc as there is no {\it a priori} reason for expanding the distribution in powers of the particle momenta, whereas the Chapman-Enskog method is expected to work only near equilibrium. In fact, for both Grad and CE approaches, $\delta f$ becomes negative at large momenta and is unable to describe substantial deviations from local isotropy. These distributions are thus unsuitable for describing the very early stages of heavy-ion collisions, where the medium expands rapidly along the longitudinal (or beam) direction of the colliding nuclei. A form of the distribution function that does not become negative at any momenta and can handle such large deviations from local momentum isotropy is given by the Romatschke-Strickland ansatz \cite{Romatschke:2003ms, Tinti:2015xwa}: 
\begin{align}\label{f_aHydro}
    f_{RS} = \exp \left( - \beta_{RS} \, \sqrt{p_\mu \, p_\nu \Xi^{\mu\nu}} \right). 
\end{align}
In the above, $\Xi^{\mu\nu} = u^\mu u^\nu + \xi^{\mu\nu} - \Delta^{\mu\nu} \, \psi$, where $\beta_{RS}$ plays the role of an effective inverse temperature, $\xi^{\mu\nu}$ characterizes the deformation from momentum isotropy, and $\psi$ models the isotropic deviation from equilibrium. Anisotropic hydrodynamics (aHydro) \cite{Florkowski:2010cf, Martinez:2010sc, Martinez:2012tu, Strickland:2014pga, Florkowski:2014bba, Molnar:2016vvu, Molnar:2016gwq, Alqahtani:2017mhy, Nopoush:2019vqc} assumes that the leading part of a distribution function that may describe early-time dynamics of fluids formed in heavy-ion collisions, is given by $f \approx f_{RS}$. This assumption for $f(x,p)$, supplemented by matching conditions relating the parameters $(\beta_{RS}, \xi^{\mu\nu}, \psi)$ to energy density, shear and bulk stresses, is then used to truncate the system of Eqs. (\ref{bulk_Denicol}-\ref{shear_Denicol}). The aHydro approach can be further improved by including corrections to the Romatschke-Strickland distribution, $f \approx f_{RS} + \delta f$, where $\delta f$ is usually obtained using a 14-moment approximation. This procedure is used to formulate `viscous anisotropic hydrodynamics' \cite{Bazow:2013ifa, McNelis:2018jho}. 

\subsection{The maximum entropy approximation}
\label{sec_mehydro_intro}

A novel way of constructing a distribution function from knowledge of the hydrodynamic variables was proposed in \cite{Everett:2021ulz}. It is based on the idea that a `least biased' distribution function that uses all of and only information of hydrodynamic moments is the one that maximizes the non-equilibrium entropy density\footnote{%
    Eq. (\ref{H-function}) holds for particles obeying classical statistics. To accommodate quantum statistics, the term in the integrand $f (\ln f - 1)$ has to be generalized to $f \ln (f) - (1 + \theta f) \ln( 1 + \theta f)/\theta$, where $\theta = 1, -1$ correspond to Bose-Einstein and Fermi-Dirac particles, respectively.
},
\begin{align}\label{H-function}
    s[f] = - \int dP \, \left(u \cdot p \right) \, f \, \left( \ln{f} - 1 \right),
\end{align}
 subject to the constraints that $f$ reproduces the instantaneous values of hydrodynamic quantities,
 \begin{align}
     e &= \int dP \, \left(u \cdot p \right)^2 \, f, \nonumber \\
     P + \Pi &= - \frac{1}{3} \, \int dP \, \left( \Delta_{\mu\nu} \, p^\mu \, p^\nu \right) \, f, \nonumber \\
     \pi^{\mu\nu} &= \int dP \, p^{\langle \mu} \, p^{\nu \rangle} \, f. 
 \end{align}
 Taking the functional derivative of $s[f]$ and employing the method of Lagrange multipliers, the maximum-entropy or `least-biased' distribution for Boltzmann particles is obtained to be
\begin{align}\label{MaxEnt distribution}
    \fME = \exp\left( - \Lambda \left( u \cdot p \right) - \frac{\lambda_\Pi}{u \cdot p} p_{\langle \mu \rangle} p^{\langle \mu \rangle} - \frac{\gamma_{\langle \mu \nu \rangle}}{u \cdot p}  \, p^{\langle \mu} \, p^{\nu \rangle}  \right).
\end{align}
In the above, $\Lambda$, $\lambda_\Pi$, and $\gamma_{\langle \mu\nu\rangle}$ are Lagrange parameters corresponding to energy density, isotropic pressure, and shear stress tensor, respectively. For particles obeying quantum statistics, the maximum entropy distribution changes to $f_{ME} = [\exp(\psi) \pm 1]^{-1}$, where $+ (-)$ denotes Fermi-Dirac (Bose-Einstein) statistics, and $\psi$ is (up to a minus sign) identical to the argument of the exponential appearing in Eq. (\ref{MaxEnt distribution}).  Note that in the absence of dissipation ($\Pi = \pi^{\mu\nu} = 0$) Eq. (\ref{MaxEnt distribution}) reduces to the usual Boltzmann distribution. Also, for small deviations from local equilibrium, $\fME \approx \feq + \delta f$, with $\delta f$ approaching the Chapman-Enskog form given by Eq. (\ref{deltaf_CE}) \cite{Everett:2021ulz}. Accordingly, for small deviations from local equilibrium, the maximum-entropy truncation scheme yields second-order Chapman-Enskog hydrodynamics. 
 
It should be also noted that, similar to the aHydro ansatz (\ref{f_aHydro}), the maximum-entropy distribution is positive definite for all momenta. However, unlike the RS-distribution which is constructed precisely to match the specific symmetries associated with the initially dominating longitudinal expansion of the fluid formed in heavy-ion collisions, the maximum-entropy distribution cares only about information contained in the macroscopic currents, irrespective of the symmetries of the expansion geometry. As a result, it can be expected to describe a considerably larger class of fluid evolutions than the Romatschke-Strickland ansatz (\ref{f_aHydro}). In this work we will compare the effectiveness of ME and anisotropic hydrodynamics in describing the macroscopic collective evolution of systems whose microscopic dynamics is controlled by the RTA Boltzmann equation. We shall consider two highly symmetric flow patterns that can be regarded as idealized rough approximations for different heavy-ion evolution stages and for which the RTA Boltzmann equation can be solved exactly -- Bjorken \cite{Bjorken:1982qr} and Gubser \cite{Gubser:2010ui, Gubser:2010ze} flow. 

We close this Section with a brief comparative discussion of our ME approach \cite{Everett:2021ulz}  with the DTT (Dissipative Type Theory) approach proposed in \cite{Calzetta:2019dfr}. For a conformal system, the authors of \cite{Calzetta:2019dfr} try to obtain a distribution (referred to as $f_\mathrm{DTT}$) that locally maximizes the \textit{entropy production rate} while matching the ideal and shear viscous components of the energy-momentum tensor. As the rate of entropy production is determined by the collisional kernel, this distribution depends on the relaxation time scale $\tau_R$. This is not the case for $\fME$ which is constructed entirely from macroscopic hydrodynamic input and agnostic about the underlying microscopic kinetics. Surprisingly, after a suitable re-definition of the Lagrange parameters, the authors of \cite{Calzetta:2019dfr} arrive at a form for $f_\mathrm{DTT}$ identical to the conformal version of Eq.~(\ref{MaxEnt distribution}). Using this result to close the exact shear evolution equation (\ref{shear_Denicol}) reduces their DTT hydrodynamic framework exactly to ME hydrodynamics. Unfortunately, the last step in their derivation involves an expansion in deviations from equilibrium, and keeping only the leading term (as done in \cite{Calzetta:2019dfr} leads to an approximation for $f_\mathrm{DTT})$ that maximizes the entropy production rate if and only if the system is already in thermal equilibrium, i.e. iff $f_\mathrm{DTT}=\feq$. Hence, with their approximation $f_\mathrm{DTT} \approx\fME$, the results presented in \cite{Calzetta:2019dfr} reproduce ME hydrodynamics. Our work here extends theirs in multiple directions, most importantly for non-conformal systems.

\section{Bjorken flow}
\label{sec_Bjorken_main}

Bjorken flow \cite{Bjorken:1982qr} describes the early stages of matter evolution in ultra-relativistic heavy-ion collisions. In this description, the system is assumed to expand boost-invariantly along the $z-$ (beam or longitudinal) direction with a $z \to -z$ symmetry, while being homogeneous and rotationally invariant in the $(x-y)$-plane (transverse to the beam direction). Mathematically, these assumptions imply invariance of the system under the combined $SO(1,1) \otimes ISO(2) \otimes Z_2$ symmetry group\footnote{%
    $SO(1,1)$ for longitudinal boost-invariance, $ISO(2)$ for rotational and translational invariance in transverse plane, and $Z_2$ for reflection symmetry.
}
\cite{Gubser:2010ze, Gubser:2010ui} which, although obscure in Cartesian coordinates, becomes manifest in the Milne coordinates:
\begin{align}
    \tau &= \sqrt{t^2 - z^2}, \, r = \sqrt{x^2 + y^2}, \nonumber \\
    \phi &= \tan^{-1}(y/x), \,\, \eta = \tanh^{-1}(z/t).
\end{align}
Here $\tau$ is the longitudinal proper time and $\eta$ is the space-time rapidity. In Milne coordinates, the metric tensor takes the form $g_{\mu\nu} = \mathrm{diag}(1, -1, -r^2, -\tau^2)$ such that the line element given by
\begin{align}
    ds^2 = d\tau^2 - dr^2 - r^2 \, d\phi^2 - \tau^2 \, d\eta^2,
\end{align}
is manifestly invariant under the Bjorken symmetries mentioned above. More importantly, the fluid appears to be at rest in these coordinates, $(u^\tau = 1, u^x = u^y = u^\eta = 0)$, which is the unique flow velocity profile consistent with the combined $SO(1,1) \otimes ISO(2) \otimes Z_2$ symmetry group. These symmetries further imply that all macroscopic quantities (like temperature, shear stress tensor etc.) are functions solely of the proper time $\tau$ such that partial differential equations of hydrodynamics get replaced by ordinary ones.   

\subsection{Exact solution of the RTA Boltzmann equation in Bjorken flow}

We consider a fluid undergoing Bjorken expansion and assume that the medium is composed of particles whose dynamics is described by kinetic theory. Bjorken symmetries at the microscopic level imply that the single particle distribution function $f(x, p)$ depends on $\tau$, the transverse momenta $p_T = \sqrt{p_x^2 + p_y^2}$ and boost-invariant longitudinal momentum $p_\eta$ \cite{Baym:1984np, Florkowski:2013lya}. The Boltzmann equation with a collisional kernel in the relaxation-time approximation \cite{anderson1974relativistic} is
\begin{align}\label{RTA}
    \frac{\partial f}{\partial \tau} = - \frac{1}{\tau_R} \left( f - \feq\right),
\end{align}
with the equilibrium distribution $\feq = \exp(- p^\tau/T)$, where $p^\tau = \sqrt{p_T^2 + p_\eta^2/\tau^2 + m^2}$ is the particle energy and $T$ is the temperature. The relaxation time is parameterized as $\tau_R \equiv  5C/T$ with constant $C$. The shear stress tensor of the fluid is diagonal, $\pi^{\mu\nu} = \mathrm{diag}(0, \pi/2, \pi/2, -\pi/\tau^2)$ with a single independent degree of freedom, $\pi \equiv \pi^\eta_\eta$. Correspondingly, the energy-momentum tensor, $T^{\mu\nu} = \mathrm{diag}(e, P_T, P_T, P_L)$, has only 3 independent components, with $P_T \equiv P + \Pi + \pi/2$ and $P_L \equiv P + \Pi - \pi$ being the effective transverse and longitudinal pressures. 

The RTA Boltzmann equation (\ref{RTA}) is formally solved exactly by \cite{Baym:1984np, Florkowski:2013lya, Florkowski:2014sfa}
\begin{align}\label{RTA_sol}
    f(\tau; p_T, p_\eta) &= D(\tau, \tau_0) \, f_0(p_T, p_\eta) 
\\\nonumber
    & + \int_{\tau_0}^{\tau} \, \frac{d\tau'}{\tau_R(\tau')} \, D(\tau, \tau') \, \exp\bigl(-p^\tau(\tau')/T(\tau') \bigr),
\end{align}
where $f_0(p_T, p_\eta)$ is the initial distribution function, and the `damping function' $D(\tau_2, \tau_1)$ is defined as
\begin{align}
D(\tau_2, \tau_1) \equiv \exp\left(- \int_{\tau_1}^{\tau_2} \, \frac{d\tau'}{ \tau_R(\tau')} \right).    
\end{align}
In this work we consider an initial distribution of generalized Romatschke-Strickland form \cite{Chattopadhyay:2021ive, Jaiswal:2021uvv}:
\begin{align}\label{f_in}
    \!\!\!\!
    f_0(p_T, p_\eta) =  \exp\left(\alpha_0 -  \frac{\sqrt{p_T^2 + (1{+}\xi_0) p_\eta^2/\tau_0^2 + m^2}}{T^{RS}_0} \right).
\end{align}
Here $T^{RS}_0$ sets the momentum scale, $\xi_0$ parametrizes the momentum-space anisotropy, and the fugacity-like parameter $\alpha_0$ models a distribution that differs from $\feq$ by a multiplicative factor even if the anisotropy $\xi_0$ is chosen to vanish. These three parameters can be tuned to generate all possible values for the three independent components of the energy momentum tensor \cite{Chattopadhyay:2021ive, Jaiswal:2021uvv}.

In order to explicitly determine the solution $f(\tau; p_T, p_\eta)$ in Eq. (\ref{RTA_sol}) one needs the time evolution of the temperature. This is obtained by imposing the energy conservation condition through Landau matching:
$e(\tau) = e_\mathrm{eq}(T(\tau))$. 
The solution for temperature is then determined from an integral equation \cite{Banerjee:1989by, Florkowski:2013lya, Florkowski:2014sfa}:
\begin{align}
    & e_\mathrm{eq}(T(\tau)) = D(\tau, \tau_0) \, \frac{ (T^{RS}_0)^4 \, e^{\alpha_0}}{4\pi^2} \, \tilde{H}_e \Bigl(\frac{\tau_0}{\sqrt{1+\xi_0} \tau}, \frac{m}{T^{RS}_0}  \Bigr) \nonumber \\
    & + \frac{1}{4\pi^2} \, \int_{\tau_0}^{\tau} \, \frac{d\tau'}{\tau_R(\tau')} \, D(\tau,\tau') \, T^4(\tau') \, \tilde{H}_e \Bigl(\frac{\tau'}{\tau}, \frac{m}{T(\tau')} \Bigr). \label{T_sol_RTA}
\end{align}
In the above the temperature dependence of $e_\mathrm{eq}(T)$ is specified by Eq. (\ref{e_eq}), and the function $\tilde{H}_e[y,z]$ is defined as,
\begin{align}
    \tilde{H}[y,z] = \int_0^\infty \, du \, u^3 \, \exp\left( - \sqrt{y^2 + z^2} \right) \, H_e\left(y, \frac{z}{u} \right),
\end{align}
with \cite{Florkowski:2014sfa}
\begin{align}
    H_e(y,z) = y \left(\sqrt{y^2 + z^2} + \frac{1+z^2}{\sqrt{y^2-1}} \, \tanh^{-1}\sqrt{\frac{y^2-1}{y^2+z^2}} \right).
\end{align}
Eq. (\ref{T_sol_RTA}) is solved for $T(\tau)$ by numerical iteration. The solution for $T(\tau)$ can then be used in Eq. (\ref{RTA_sol}) to obtain the distribution function at any time. One may also directly use the following formulae to calculate the effective transverse and longitudinal pressures,
\begin{align}
\label{PT_sol_RTA}
    P_T(\tau) =&  D(\tau,\tau_0) \frac{(T^{RS}_0)^4}{8 \pi^2 \alpha_0} \tilde{H}_{T} \left(\frac{\tau_0}{\tau \sqrt{1+\xi_0}}, \frac{m}{T^{RS}_0} \right) 
\\\nonumber 
    & + \frac{1}{8\pi^2}  \int_{\tau_0}^{\tau} \frac{d\tau'}{\tau_R(\tau')} D(\tau,\tau') T^4(\tau') \tilde{H}_{T} \left(\frac{\tau'}{\tau},\frac{m}{T(\tau')}\right), 
\\
\label{PL_sol_RTA}
    P_L(\tau) =&  D(\tau,\tau_0) \frac{(T^{RS}_0)^4}{4 \pi^2 \alpha_0} \tilde{H}_{L} \left(\frac{\tau_0}{\tau \sqrt{1+\xi_0}}, \frac{m}{T^{RS}_0} \right) 
\\\nonumber  
    & + \frac{1}{4\pi^2}  \int_{\tau_0}^{\tau} \frac{d\tau'}{\tau_R(\tau')} D(\tau,\tau') T^4(\tau') \tilde{H}_{L} \left(\frac{\tau'}{\tau},\frac{m}{T(\tau')}\right). 
\end{align}
Here the functions $\tilde{H}_{T,L}$ are defined by \cite{Florkowski:2014sfa}
\begin{align}
\nonumber
    \tilde{H}_{T,L}(y,z) &\equiv \int_{0}^{\infty} du \,  u^3  \exp\Bigl(-\sqrt{u^2 + z^2}\Bigr)\, H_{T,L}\left(y,\frac{z}{u}\right), 
\end{align}
with
\begin{align}
    H_{T}(y,z) =&  \frac{y}{(y^2-1)^{3/2}} \bigg[ - \sqrt{(y^2-1)(y^2+z^2)} 
\nonumber \\    
    & + \left( z^2 + 2y^2 - 1 \right) \tanh^{-1}\sqrt{\frac{y^2-1}{y^2+z^2}} \bigg] ,
\end{align}
\begin{align}
    H_{L}(y,z) =& \frac{y^3}{(y^2-1)^{3/2}} \bigg[ \sqrt{(y^2-1)(y^2+z^2)}  
\nonumber \\
    & -  \left( z^2 + 1 \right) \tanh^{-1}\sqrt{\frac{y^2-1}{y^2+z^2}} \bigg].
\end{align}
From Eqs.~(\ref{PT_sol_RTA},\ref{PL_sol_RTA}) it is straightforward to obtain the bulk and shear viscous stresses as $\Pi = \frac{1}{3}(P_L + 2 P_T - 3 P)$ and $\pi = \frac{2}{3}(P_T {-} P_L)$.

\subsection{Maximum entropy truncation for Bjorken flow}
\label{secIIB}

We now describe the maximum entropy truncation scheme \cite{Everett:2021ulz} for the case of Bjorken flow. Using the RTA Boltzmann equation (Eq. (\ref{RTA})), one may derive exact evolution equations for the three independent components of energy-momentum tensor, viz., $e, P_L, P_T$:
\begin{align} 
    \frac{de}{d\tau} &= - \frac{1}{\tau} \, \left( e + P_L \right), \label{e_evol}\\
    \frac{dP_L}{d\tau} &= - \frac{P_L - P}{\tau_R} + \frac{\zeta^{L}_{z}}{\tau}, \label{PL_evol}\\
    \frac{dP_T}{d\tau} &= - \frac{P_T - P}{\tau_R} + \frac{\zeta^{\perp}_{z}}{\tau}, \label{PT_evol} 
\end{align}
where 
\begin{align}
\label{zeta_long}
   \zeta^{L}_{z} = - 3 P_L + I^{\mathrm{exact}}_{240}, \,\,\,\,
   \zeta^{\perp}_{z} = - P_T + I^{\mathrm{exact}}_{221},
\end{align}
using the definition
\begin{equation}
\label{I_exact}
    I^{\mathrm{exact}}_{nrq} = \frac{1}{(2q)!!} \int dP \, (p^\tau)^{n-r-2q} \, \left(\frac{p_\eta}{\tau}\right)^r \, p_T^{2q} \, f
\end{equation}
with $dP \equiv d^2p_{T} dp_\eta/[(2\pi)^3 \tau p^\tau]$. 

Equations (\ref{e_evol}-\ref{PT_evol}) are exact but not closed, owing to the couplings $\zeta^{L}_{z}$ and $\zeta^{\perp}_{z}$ which require knowledge of the solution of the Boltzmann equation (\ref{RTA}). In the maximum entropy truncation scheme one evaluates the moments $I_{240}$ and $I_{221}$ in Eq.~(\ref{zeta_long}) approximately by replacing $f$ in Eq.~(\ref{I_exact}) by the maximum entropy distribution $\fME$. In Bjorken coordinates Eq.~(\ref{MaxEnt distribution}) takes the form
\begin{equation}
\label{f_ME_Bjorken}
   \fME = \exp\Bigl(- \Lambda p^\tau - \frac{\lambda_\Pi}{p^\tau} \left( p_T^2 + p_\eta^2/\tau^2 \right)
   - \frac{\gamma_{ij} p^i p^j}{p^\tau} \Bigr),
\end{equation}
where the indices $\{i,j\}$ run over $\{x,y,\eta\}$. The assumption $f \approx \fME$ closes the system of equations (\ref{e_evol}-\ref{PT_evol}). Due to the symmetries of Bjorken flow the traceless tensor $\gamma_{ij}$ in Eq.~(\ref{f_ME_Bjorken}) becomes diagonal, with a single independent component: $\gamma_{ij} = \mathrm{diag} (0,  \gamma/2, \gamma/2, - \tau^2 \gamma)$. Accordingly, the scalar $\gamma_{ij} p^i p^j$ simplifies to $\gamma \, (p_T^2/2 - p_\eta^2/\tau^2)$. The three Lagrange parameters $(\Lambda, \lambda_\Pi, \gamma)$ have to be chosen such that $\fME$ reproduces the three independent components of $T^{\mu\nu}$ at each instant of time (matching conditions):
\begin{equation}\label{e_PL_PT_matching_conditions}
    e = \tilde{I}_{200}, \qquad P_L = \tilde{I}_{220}, \qquad P_T = \tilde{I}_{201}.
\end{equation}
Here $\tilde{I}_{nrq}$ denotes moments of the maximum entropy distribution:
\begin{equation}
\label{I_ME}
    \tilde{I}_{nrq} = \frac{1}{(2q)!!} \int dP \, (p^\tau)^{n-r-2q} \, \left(\frac{p_\eta}{\tau}\right)^r \, p_T^{2q} \, \fME.
\end{equation}
Thus, instead of Eqs.~(\ref{e_evol}-\ref{PT_evol}), we shall solve
\begin{align} 
    \frac{de}{d\tau} &= -\frac{e + P_L}{\tau}, 
\label{e_evol_ME}
\\
    \frac{dP_L}{d\tau} &= - \frac{P_L - P}{\tau_R} + \frac{\tilde{\zeta}^{L}_{z}}{\tau},    
\label{PL_evol_ME}
\\
    \frac{dP_T}{d\tau} &= - \frac{P_T - P}{\tau_R} + \frac{\tilde{\zeta}^{\perp}_{z}}{\tau},
\label{PT_evol_ME} 
\end{align}
with
\begin{align}
\label{tilde_zeta}
   \tilde{\zeta}^{L}_{z} = - 3 P_L + \tilde{I}_{240}, \,\,\,\,
   \tilde{\zeta}^{\perp}_{z} = - P_T + \tilde{I}_{221}.
\end{align}
For brevity we will refer to Eqs.~(\ref{e_evol_ME})-(\ref{PT_evol_ME}) as {\it Maximum Entropy (ME) hydrodynamics} for Bjorken flow.

Instead of directly solving these equations, which at every time step involves a 3-dimensional inversion to obtain from $(e,\,P_L,\,P_T)$ the Lagrange multipliers $(\lambda,\lambda_\Pi,\gamma)$ needed for evaluating the couplings $(\tilde{\zeta}^L_z,\tilde{\zeta}^{\perp}_z)$, we shall recast Eqs.~(\ref{e_evol_ME}-\ref{PT_evol_ME}) as evolution equations of the Lagrange parameters themselves: Defining $X^a \equiv \{e,P_L,P_T \}$ and $x^a \equiv \{\Lambda, \lambda_\Pi, \gamma \}$, we write
\begin{align}
    dX^a & = M^{a}_{\,\,\,b} \,  dx^b,
\end{align}    
with $M^{a}_{\,\,\,b} \equiv \partial X^a/\partial x^b$, and invert this to obtain 
\begin{align}
    \frac{dx^a}{d\tau} &= \left( M^{-1} \right)^{a}_{\,\,\,b} \, \frac{dX^b}{d\tau}
\end{align}
as evolution equations for the Lagrange multipliers. Using local rest frame coordinates for simplicity, $p_{z,\mathrm{LRF}}{\,=\,}p_\eta/\tau$ such that $E_{\mathrm{LRF}} = \sqrt{p^2_{\mathrm{LRF}}{+}m^2}$ with $p^2_{\mathrm{LRF}} = p_T^2 + p^2_{z,\mathrm{LRF}}$, and dropping the LRF subscript to ease clutter, the maximum entropy distribution reads\footnote{
    We note in passing that for $\gamma{\,=\,}0$ the maximum entropy distribution (\ref{fME_LRF}) is isotropic in the LRF, and hence $\gamma{\,=\,}0$ is equivalent to zero shear stress, $\pi{\,=\,}0$.
}
\begin{equation}
\label{fME_LRF}
   \fME = \exp\Bigl(-\Lambda E_p -  \frac{\lambda_\Pi}{E_p} \, p^2
    - \frac{\gamma}{E_p} \left( p_T^2/2 - p_z^2 \right) \Bigr).
\end{equation}
Starting from the matching conditions (\ref{e_PL_PT_matching_conditions}), 
\begin{align}
    e &= \int dP \, E_p^2 \, \fME, \quad
\\ \nonumber
    P_{L} &= \int dP \, p_{z}^2 \, \fME, \quad 
    P_{T}  = \frac{1}{2} \int dP \, p_{T}^2 \, \fME,
\end{align}
with $dP= d^3p/[(2\pi)^3 E_p]$, the first row of the matrix $M$ is found to have the elements
\begin{align} \label{M_row1}
    M^{e}_{\,\, \Lambda} & \equiv \frac{\partial e}{\partial \Lambda} = -\! \int dP \, E_p^3 \, \fME = - \tilde{I}_{300},
\nonumber\\ \!\!\!\!
    M^{e}_{\,\, \lambda_\Pi} & \equiv \frac{\partial e}{\partial \lambda_\Pi} = -\! \int dP \, E_p \, p^2 \, \fME = - 2 \tilde{I}_{301} - \tilde{I}_{320}, 
\\\nonumber
    M^{e}_{\,\, \gamma} &\equiv \frac{\partial e}{\partial \gamma} = - \int dP \,E_p \, \left(p_T^2/2 - p_z^2 \right) = - \tilde{I}_{301} + \tilde{I}_{320}.
\end{align}
Similarly, the second row has the elements
\begin{align}\label{M_row2}
    M^{P_L}_{\,\, \Lambda} &\equiv \frac{\partial P_L}{\partial \Lambda} = - \int dP \, E_p \, p_z^2 \, \fME = - \tilde{I}_{320},
\nonumber\\
    M^{P_L}_{\,\, \lambda_\Pi} & \equiv \frac{\partial P_L}{\partial \lambda_\Pi} = - \int dP \, E_p^{-1} \, p_z^2 \, p^2 \, \fME \\
    & = - 2 \tilde{I}_{321} - \tilde{I}_{340}, 
\nonumber \\\nonumber
    M^{P_L}_{\,\, \gamma}& \equiv \frac{\partial P_L}{\partial \gamma} = - \int dP \,E_p^{-1} \, p_z^2 \, \left(p_T^2/2 - p_z^2 \right) 
\\\nonumber
    & = - \tilde{I}_{321} + \tilde{I}_{340}.
\end{align}
The components of the third row are
\begin{align}\label{M_row3}
    M^{P_T}_{\,\, \Lambda} &\equiv \frac{\partial P_T}{\partial \Lambda} = - \frac{1}{2} \int dP \, E_p \, p_T^2 \, \fME = - \tilde{I}_{301},
\nonumber\\
    M^{P_T}_{\,\, \lambda_\Pi} & \equiv \frac{\partial P_T}{\partial \lambda_\Pi} = - \frac{1}{2} \int dP \, E_p^{-1} \, p_T^2 \, p^2 \, \fME 
\\
    & = - 4 \tilde{I}_{302} - \tilde{I}_{321},
\nonumber \\ \nonumber
    M^{P_T}_{\,\, \gamma}& \equiv \frac{\partial P_T}{\partial \gamma} = - \frac{1}{2} \int dP \,E_p^{-1} \, p_T^2 \, \left(p_T^2/2 - p_z^2 \right) 
\\\nonumber
    & = - 2\tilde{I}_{302} + \tilde{I}_{321}.
\end{align}
In spherical polar coordinates $(p, \theta_p, \phi_p)$ the moments $\tilde{I}_{nrq}$ read
\begin{align}
\label{Itilde-spherical}
    \tilde{I}_{nrq} &= \frac{1}{(2q)!!} \int dP \, E_{p}^{n-r-2q} \, p^{r+2q} \, \cos^r{\theta_p} \, \sin^{2q}{\theta_p} \, 
\nonumber \\
    & \times \exp\Bigl(-\Lambda E_p - \frac{\lambda_\Pi}{E_p} p^2
   - \frac{\gamma \, p^2}{E_p} \bigl( \textstyle{\frac{1}{2}}\sin^2{\theta_p} - \cos^2{\theta_p} \bigr) \Bigr), 
\nonumber \\
   &= \frac{1}{(2q)!!} \int_0^{\infty} \frac{d p}{4\pi^2} \, E_p^{n-r-2q-1} \, p^{r+2q+2} \, 
\nonumber \\
   & \times \exp\left(-\Lambda E_p - \frac{\lambda_\Pi p^2}{E_p} \, - \frac{\gamma\, p^2}{2E_p} \right) \, R_{rq}\left(\frac{3 \gamma p^2}{2E_p} \right)
\end{align}
where 
\begin{align}
\!\!\!
    R_{rq}(\alpha) \equiv \int_{0}^{\pi}\!\! d\theta_p \, \cos^r{\theta_p} \, \sin^{2q + 1} \theta_p \, \exp\left( \alpha \, \cos^2\theta_p \right).
\end{align}
The integrals $R_{rq}(\alpha)$ can be expressed analytically in terms of error functions. Note that $\alpha = 3\gamma p^2/2E_p$ has the same sign as $\gamma$ which can be positive or negative. For $\alpha<0$ we can define $t(\alpha) = \sqrt{\pi} \, \mathrm{Erf}(\sqrt{-\alpha})/\sqrt{-\alpha}$ which is well-behaved as $\alpha \to - \infty$. Listing only the ones required in this analysis, the $R_{rq}$ functions can then be expressed as
\begin{align}
    R_{00} &= t,   \quad 
    R_{01} = - \frac{e^\alpha}{\alpha} + \frac{t (1 + 2\alpha)}{2 \alpha}, 
\nonumber \\
    R_{02} &= - \frac{e^\alpha (3+2\alpha)}{2\alpha^2} + \frac{t(3 + 4\alpha + 4\alpha^2)}{4\alpha^2},
\nonumber \\ 
    R_{20} &= \frac{e^\alpha}{\alpha} - \frac{t}{2\alpha}, \quad
    R_{21} = \frac{3 e^\alpha}{2\alpha^2} - \frac{t (3+2\alpha)}{4\alpha^2}, 
\nonumber \\
    R_{40} &= \frac{e^\alpha (2\alpha - 3)}{2 \alpha^2} + \frac{3t}{4\alpha^2}.
\label{R_alpha_negative}
\end{align} 
For positive $\alpha >0$, however, the right hand sides of Eqs.~(\ref{R_alpha_negative}) are inconvenient because they are sums of terms that individually diverge exponentially in the limit $\alpha\to\infty$. This is made explicit by writing 
\begin{align}
    t(\alpha) = \frac{\sqrt{\pi}\, \mathrm{Erf}(\sqrt{-\alpha})}{\sqrt{-\alpha}} = \frac{2 \, e^\alpha \, {\cal D}(\sqrt{\alpha})}{\sqrt{\alpha}}, \,\,\,\, \mathrm{for} \, \alpha >0, 
\end{align}
where the DawsonF function (available at \cite{faddeeva} for numerical implementation in C++) is well-behaved as $\alpha \to \infty$. While these exponential divergences all cancel between the various terms on the right hand sides of Eqs.~(\ref{R_alpha_negative}), they should be removed analytically before numerical implementation. This can be achieved by extracting a factor $e^\alpha$ from the $R_{rq}$ functions and combining it with the exponential prefactor in Eq.~(\ref{Itilde-spherical}), defining $\tilde{R}_{rq}(\alpha)=e^{-\alpha} R_{rq}(\alpha)$ which is manifestly free from exponential divergences in the limit $\alpha\to+\infty$. For the numerical implementation of the moments $\tilde{I}_{nrq}$ we therefore use the following expressions:
\begin{align}\label{tilde_Inrq}
    \tilde{I}_{nrq} &= \frac{1}{(2q)!!} \int_0^{\infty} \frac{d p}{4\pi^2} \, E_p^{n-r-2q-1} \, p^{r+2q+2} \, R_{rq}\left(\frac{3 \gamma p^2}{2E_p} \right) 
\nonumber \\
    & \times \exp\left(-\Lambda E_p - \frac{\lambda_\Pi p^2}{E_p} \, - \frac{\gamma\, p^2}{2E_p} \right),  \,\,\,\,\, \gamma<0 
\nonumber \\
    \tilde{I}_{nrq} &= \frac{1}{(2q)!!} \int_0^{\infty} \frac{d p}{4\pi^2} \, E_p^{n-r-2q-1} \, p^{r+2q+2} \, \tilde{R}_{rq}\left(\frac{3 \gamma p^2}{2E_p} \right) 
\nonumber \\
    & \times \exp\left(-\Lambda E_p - \frac{\lambda_\Pi p^2}{E_p} \, + \frac{\gamma\, p^2}{E_p} \right),  \,\,\,\,\, \gamma>0.
\end{align}
Note that, since the functions $R$ and $\tilde{R}$ are well-behaved in the regions where they are used in Eqs.~(\ref{tilde_Inrq}), convergence of these integrals demands that the exponential factors in Eqs.~(\ref{tilde_Inrq}) fall off at large momenta. This requires
\begin{equation}\label{bound_Bjorken}
    \Lambda + \lambda_\Pi > \bigg|\mathrm{min}\Bigl(\frac{\gamma}{2}, -\gamma \Bigr)\bigg|.
\end{equation}
This criterium was already deduced in \cite{Everett:2021ulz} where it was found that $\Lambda + \lambda_\Pi > \big|\mathrm{min}(\gamma_1, \gamma_2, \gamma_3)\big|$ where $\gamma_i$ are the eigenvalues of the shear stress tensor in the fluid rest frame. For Bjorken flow, Milne coordinates are the local rest frame coordinates, and thus $\gamma_1 = \gamma_2 = \gamma/2$ and $\gamma_3 = - \gamma$. We note that although the Lagrange multiplier $\Lambda$ appears similar to an inverse temperature $\beta$ it does not need to be positive for $\fME$ to be well-behaved; all that is required is the sum $\Lambda + \lambda_\Pi$ being positive. This feature will manifest itself in Sec. \ref{non-conformal dynamics} where we generate negative bulk viscous pressures using $\fME$. 

\subsection{Note on initial conditions} \label{sec:ic}

\begin{figure}
\includegraphics[width = 0.95\linewidth]{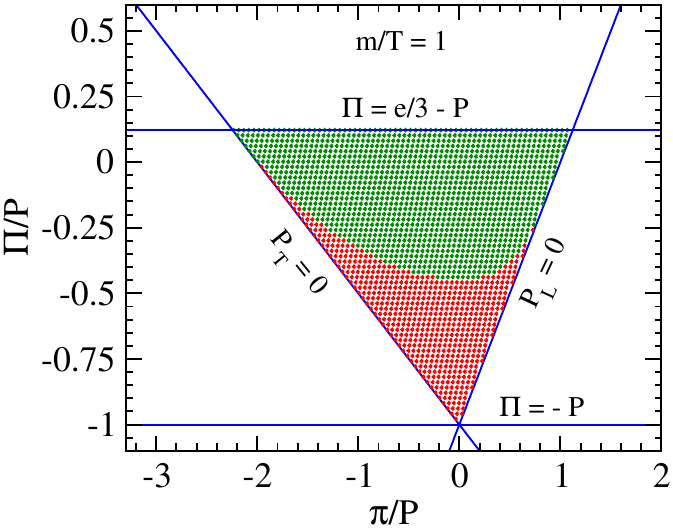}
\caption{Shear and bulk stresses, scaled by the thermal pressure $P$, generated with $\fME$ for different values of the Lagrange multipliers at fixed $m/T = 1$. The red points indicate unphysical regions of negative Boltzmann entropy.
\vspace*{-3mm}} 
\label{fig:triangle_MB}
\end{figure}

In this subsection we explore the range of bulk and shear stresses that can be accessed with the maximum-entropy ansatz (\ref{fME_LRF}) for the distribution function, as well as its quantum statistical generalization. For Bjorken flow, positivity of the distribution function $f(x,p)$ implies that the effective longitudinal and transverse pressures, $P_L$ and $P_T$, are both positive. Also, for non-zero particle mass $m$ the trace of the energy-momentum tensor is non-negative. These constraints imply that the bulk and shear stresses (in units of the thermal pressure) satisfy the following inequalities \cite{Chattopadhyay:2021ive}:
\begin{align}\label{bound_pi_PI}
    \tilde{\pi} - \tilde{\Pi} < 1, \,\, \frac{\tilde{\pi}}{2} + \tilde{\Pi} > -1, \,\,
  \tilde{\Pi} \leq \frac{\tilde{e}}{3} - 1,
\end{align}
where $\tilde{A} \equiv A/P$. Equations~(\ref{bound_pi_PI}) restrict the dissipative fluxes to lie within a triangular region in the scaled shear and bulk pressure plane, as depicted in Fig.~\ref{fig:triangle_MB}. Note that the upper bound on $\tilde{\Pi}$ depends on the mass of the constituent particles, and for Fig.~\ref{fig:triangle_MB} we chose $m/T = 1$.\footnote{%
    For conformal systems the allowed region in Fig.~\ref{fig:triangle_MB} shrinks to the line $\Pi = 0$, and the scaled shear stress can vary from $-2$ to $1$. 
}

Let us consider the lower part of this triangular region, where the shear stress is small and the scaled bulk viscous pressure is negative and large in magnitude. The limit $\tilde{\Pi}{\,=\,}{-1},\, \tilde{\pi}{\,=\,}0$ characterises a state where the effective pressures vanish: $P_L{\,=\,}P_T{\,=\,}0$ \cite{Chattopadhyay:2021ive, Jaiswal:2021uvv}. More specifically, as the temperature is held fixed at a value comparable to the particle mass, the state requires $f(x,p) \sim {\cal A} \, \delta(p)/p^2$ such that the energy density stems entirely from the rest mass of the particles. The modified Romatschke-Strickland distribution (\ref{f_in}) can accommodate such extreme states, due to the fugacity factor $\alpha_0$ which provides control over the normalization factor ${\cal A}$. States with $P_L = P_T = 0$ (i.e. $\tilde{\Pi}{\,=\,}-1,\,\tilde{\pi}{\,=\,}0$) can also be generated with the maximum-entropy distribution, by making the Lagrange parameter $\Lambda$ sufficiently negative, as discussed in Appendix~A of \cite{Jaiswal:2021uvv}.\footnote{%
    For $\Lambda < 0$, the enhancement at low momenta stems from $\fME(p{\,=\,}0) = \exp(-\Lambda m)$.
}

However, the need for overpopulating low-momentum modes in order to generate states with large negative bulk pressures implies that classical statistics must break down in the lower part of the triangle in Fig.~\ref{fig:triangle_MB}. A physical quantity that signals this breakdown is the Boltzmann entropy (\ref{H-function}) which goes negative for distribution functions that generate $(\tilde{\pi}, \tilde{\Pi})$ pairs in the lower part of the triangle. Using $\fME$ as given in Eq.~(\ref{fME_LRF}), the non-equilibrium entropy density can be expressed in terms of macroscopic quantities as
\begin{align}\label{entropy_Boltzmann}
    s = \Lambda \, e + \lambda_\Pi \, (P_L{+}2 P_T) + \gamma \, (P_T{-}P_L) + n,
\end{align}
where $n$ is the number density of particles,\footnote{%
    Note that away from thermal equilibrium $n$ does {\it not} satisfy the equilibrium relation $n_\mathrm{eq} = P/T$.
}
\begin{align}
    n = \int dP \, E_p \, \fME\, .
\end{align}
We scan through the allowed region of phase space for $(\tilde{\pi}, \tilde{\Pi})$ with fixed $\tilde{e}$ and obtain the corresponding Lagrange parameters $(\Lambda, \lambda_\Pi, \gamma)$. The Boltzmann entropy density for these points is then computed, and depending on whether the entropy is positive or negative, we mark their positions $(\tilde{\pi}, \tilde{\Pi})$ in the triangle by green or red dots. Fig.~\ref{fig:triangle_MB} shows that for systems with $m/T \sim 1$ initial conditions for dissipative fluxes generated by $\fME$ that lie in the lower parts of the kinetically allowed triangle are not appropriately described using Boltzmann statistics.

\begin{figure}
\includegraphics[width=0.95 \linewidth]{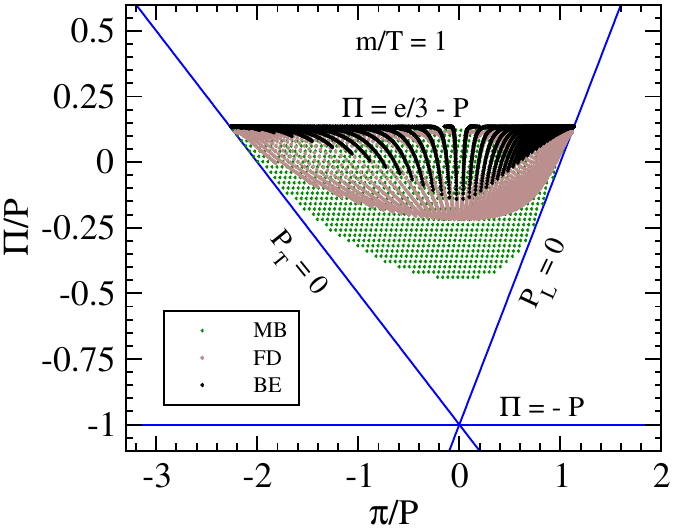}
\caption{Shear and bulk stresses generated by $\fME$ at fixed $m/T = 1$ using Maxwell-Boltzmann (MB, green), Fermi-Dirac (FD, brown), and Bose-Einstein (BE, black) statistics. For classical (MB) statistics only points with positive entropy density are shown. 
\vspace*{-3mm}}
\label{fig:triangle_MB_FD_BE}
\end{figure} 

It is, in fact, reasonable to doubt the applicability Boltzmann statistics even near the edges of the green region in Fig.~\ref{fig:triangle_MB} where the entropy density becomes small and classical statistics likely begins to break down. In Fig.~\ref{fig:triangle_MB_FD_BE} we therefore explore the ranges of $(\tilde{\pi}, \tilde{\Pi})$ that can be accessed using the maximum-entropy distributions for quantum statistics. For Bjorken flow, $\fME$ generalizes for particles with arbitrary statistic to \cite{Everett:2021ulz}
\begin{align}\label{MaxEnt distribution quantum}
   \fME = \left[\exp\Bigl(\Lambda E_p +  \frac{\lambda_\Pi}{E_p} \, p^2
    + \frac{\gamma}{E_p} \left( p_T^2/2 - p_z^2 \right) \Bigr) + \theta \right]^{-1},
\end{align}
where $\theta{\,=\,}1$ ($-1$) for Fermi-Dirac (Bose-Einstein) statistics, respectively.\footnote{%
    This does not allow for the possibility of Bose condensation which will be explored elsewhere.
} 
To generate Fig.~\ref{fig:triangle_MB_FD_BE} we scanned a wide range of values for two of the three Lagrange parameters, namely $\Lambda$ and $\gamma$, and root-solved for $\lambda_\Pi$ such that $m/T$ stays fixed at unity (for comparison with the Maxwell-Boltzmann case studied in Fig.~\ref{fig:triangle_MB}). With these Lagrange parameters the scaled fluxes are calculated for FD (brown) and BE (black) statistics, shown as scatter plots in Fig.~\ref{fig:triangle_MB_FD_BE} where they are overlaid over the (positive entropy density) green points for classical MB statistics from Fig.~\ref{fig:triangle_MB}.\footnote{%
    One shows easily that the extension of Eq.~(\ref{H-function}) to quantum statistics always yields positive values for the entropy density.
}
One sees that, once effects of quantum statistics are consistently incorporated, the fraction of $(\tilde{\pi}, \tilde{\Pi})$-space that can be accessed with the maximum-entropy parametrization of the distribution function is further reduced.\footnote{%
    We note that the accessible region for quantum statistics, as well as the region with positive entropy density for Maxwell-Boltzmann statistics, shrink when $m/T$ is reduced, and grows to cover almost the entire triangle when $m/T$ is large, $m/T>10$.} 

For computational economy we will continue to use the Maxwell-Boltzmann form (\ref{f_ME_Bjorken},\ref{fME_LRF}) of the Maximum Entropy distribution in the rest of the paper. However, in later sections of this paper dealing with non-conformal dynamics we shall restrict ourselves to initial conditions that do not lie outside the region allowed by Fermi-Dirac statistics. This guarantees positive Boltzmann entropy density in the initial state and, due to the H-theorem stating that entropy can never decrease, also at all later times.      

\subsection{Conformal dynamics} 
\label{conformal dynamics}

Before comparing results of the maximum entropy truncation scheme with exact solutions of the RTA Boltzmann equation and results from other hydrodynamic approximations for the general case of massive particles, we first study the somewhat simpler massless case. For such a conformal system the bulk viscous pressure vanishes and the energy-momentum tensor $T^{\mu\nu}{\,=\,}\mathrm{diag}\,(e, P_T, P_T, P_L)$ becomes traceless ($T^\mu_\mu = 0$). As a result, $T^{\mu\nu}$ has only two independent components for which we take the energy density $e$ and effective longitudinal pressure $P_L$. 

To obtain the evolution of $e$ and $P_L$ in the microscopic theory we solve the RTA Boltzmann equation using a standard RS-ansatz as initial condition, obtained from Eq.~(\ref{f_in}) by setting the parameters $\alpha_0$ and $m$ to zero:
\begin{align}
    f_{0} = \exp\left(- \frac{\sqrt{p_T^2  + (1{+}\xi_0) p_\eta^2/\tau_0^2 }}{T^{RS}_0} \right).
\end{align}
The conformal ($m=0$) limit of Eq.~(\ref{T_sol_RTA}) yields for the exact evolution $e(\tau)$ of the equilibrium energy density
\begin{align}
    & e_\mathrm{eq}(T(\tau)) = D(\tau, \tau_0) \frac{3 (T^{RS}_0)^4}{\pi^2} \, {\cal H}_e\left(\frac{\tau_0}{\tau \sqrt{1+\xi_0}}\right) \nonumber \\
    & + \int_{\tau_0}^{\tau} \, \frac{d\tau'}{\tau_R(\tau')} \, D(\tau, \tau') \, {\cal H}_e\left(\frac{\tau'}{\tau} \right) \, e_\mathrm{eq}(T(\tau'))
\end{align}
where 
\begin{align}\label{H_e}
    \!\!\!\!
    {\cal H}_e(x) \equiv \frac{1}{2} H_e(x,0) =  \frac{1}{2} \, \left( x^2 + \frac{\tanh^{-1}\sqrt{1{-}\frac{1}{x^2}}}{\sqrt{1{-}\frac{1}{x^2}}} \right).
\end{align}
From this the exact temperature evolution $T(\tau)$ is obtained through the equation of state $e_\mathrm{eq}(T) = 3 P = 3T^4/\pi^2$. For the relaxation time we use the conformal ansatz
\begin{align}\label{conf_tau}
    \tau_R(\tau) = 5\,\frac{C}{T(\tau)},
\end{align}
fixing $C = 10/(4\pi)$ in this subsection.\footnote{%
    For a conformal systems the parameter $C$ (which controls the interaction strength among the microscopic constituents) is equal to the specific shear viscosity, $C = \eta/s$, where $\eta$ is the shear viscosity and $s = (e{+}P)/T$ the entropy density.}
By tuning the parameters $(T^{RS}_0, \xi_0)$ we generated a variety of initial values for the normalised shear stress $\bar{\pi} \equiv \pi/(4P)$ (note that $\pi = P - P_L$) while keeping the initial temperature fixed at $T_0 = 500$ MeV. Table~\ref{table:IC_conformal_aH} tabulates the different initial values of $(T^{RS}_0, \xi_0)$ used in our analysis, along with the corresponding initial values for $\bar\pi$ and the color coding used for the evolution trajectories plotted in Fig.~\ref{fig:pibar_conformal}:  
%
\begin{table}[h!]
 \begin{center}
  \begin{tabular}{|c|c|c|c|c|c|}
   \hline
   & Blue & Green & Magenta & Maroon & Orange \\ 
   \hline
   $\bar{\pi}_0$  &  $-0.45$ &  $-0.25$  &  $0$ &  $0.15$ &  $0.245$  \\
   \hline
   \hline
    $T^{RS}_0/T_0$ & $0.411$ & $0.719$ & $1.0$ & $1.245$ & $1.959$ \\
   \hline
    $\xi_0$ & $-0.985$ & $-0.828$ & $0$ & $3.206$ & $133.595$ \\
   \hline
  \end{tabular}
  \caption{Association of initial conditions $\bar\pi \equiv \pi/(4P)$ with the Romatschke-Strickland parameters $(T^{RS}_0, \xi_0)$ used to initialise the RTA Boltzmann equation.}
\label{table:IC_conformal_aH}
 \end{center}
 \vspace*{-.6cm}
\end{table}

In the conformal limit the ME hydrodynamic equations (\ref{e_evol_ME})-(\ref{PT_evol_ME}) reduce to
\begin{align} 
    \frac{de}{d\tau} & = -\frac{e + P_L}{\tau}, 
\label{e_evol_BJ_conf}
\\
    \frac{dP_L}{d\tau} &= - \frac{P_L - P}{\tau_R} + \frac{\tilde{\zeta}^{L}_{z}}{\tau}, 
\label{PL_evol_BJ_conf}
\end{align}
where all moments are to be calculated as described in Sec.~\ref{secIIB}, albeit the particle mass $m$ set to zero. Also, we shall drop the Lagrange parameter $\lambda_\Pi$ in the maximum entropy distribution (\ref{fME_LRF}) which was introduced to match it to the bulk viscous pressure $\Pi$ which vanishes in conformal systems. To re-write Eqs.~(\ref{e_evol_BJ_conf}-\ref{PL_evol_BJ_conf}) for $X^a \equiv (e, P_L)$ in terms of the Lagrange parameters $x^a \equiv (\Lambda, \gamma)$ we use $dX^a = M^{a}_{\,\,b} \, dx^b$, which takes the explicit form 
\begin{align}\label{invert_e_PL_conformal}
    \begin{pmatrix}
         de \\ 
         d P_L
    \end{pmatrix} = 
\begin{pmatrix}
     -\tilde{I}_{300} &\ - \tilde{I}_{301}{+}\tilde{I}_{320}  \\
    - \tilde{I}_{320} &\  -\tilde{I}_{321}{+}\tilde{I}_{340} \\ 
\end{pmatrix}
\begin{pmatrix}
     d\Lambda \\ 
     d\gamma
\end{pmatrix}.
\end{align}
Calculating $\tilde{I}_{nrq}$ from Eq.~(\ref{tilde_Inrq}) with $m{\,=\,}0$, the integral over $p$ can be done analytically\footnote{%
    For any $(n, r, q)$, the integration over the variable $t$ in Eq.~(\ref{Inrq_conformal}) can also be performed analytically. However, the resulting expressions are cumbersome and therefore not listed here. For analytical expressions of $\tilde{I}_{nrq}$ in terms of a generating functional please refer to Appendix B of \cite{Calzetta:2019dfr}.  
}
such that (using $\bar{\gamma}{\,\equiv\,} \gamma/\Lambda$)
\begin{align}\label{Inrq_conformal}
    \tilde{I}_{nrq} = \frac{(n+1)!}{(2q)!! \Lambda^{n+2}} \int_{0}^{1} \frac{dt}{2\pi^2} \, \frac{\left(1 - t^2\right)^q \, t^r}{\left[ 1 + \frac{\bar{\gamma}}{2} \left( 1 - 3 t^2 \right) \right]^{n+2}}.
\end{align}
The evolution equations for $(\Lambda, \gamma)$ are obtained from $dx^a/d\tau = (M^{-1})^{a}_{\,\,b} \, dX^b/d\tau$. To match the initial ME distribution to the assumed initial temperature $T_0 = 500$\,MeV and the selected initial shear stress values listed in Tables~\ref{table:IC_conformal_aH} and \ref{table:IC_conformal_ME}, the following initial values $(\Lambda_0, \gamma_0)$ must be chosen for the Lagrange parameters:
%
\begin{figure}[!t]
\includegraphics[width=0.95\linewidth]{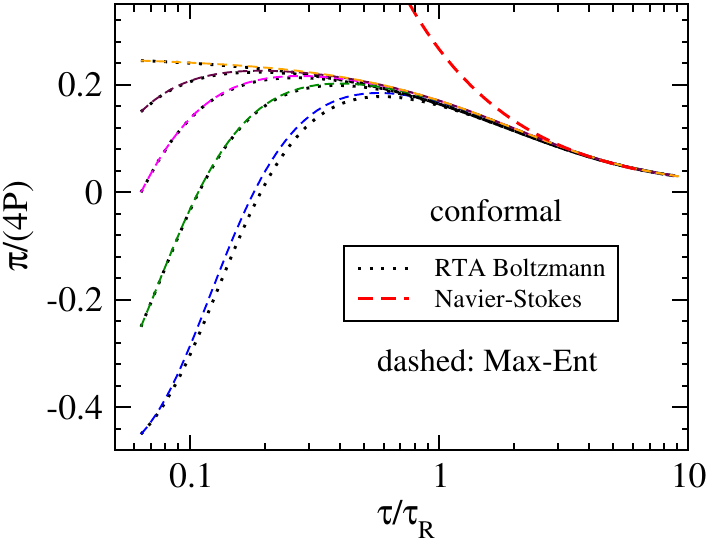}
\caption{
    Evolution of the shear inverse Reynolds number $\pi/(4P)$ from the exact solution of the RTA Boltzmann equation for a conformal gas of massless particles (dotted lines), compared with that from ME hydrodynamics (``Max-Ent'', thin dashed lines). Hydrodynamic evolution according to the first-order Navier-Stokes limit is shown as a thick red dashed line for comparison. Please refer to Tables~\ref{table:IC_conformal_aH} or \ref{table:IC_conformal_ME} for the color coding.
    \vspace*{-3mm}}
    \label{fig:pibar_conformal}
\end{figure}
%
\begin{table}[h!]
 \begin{center}
 \resizebox{0.8\columnwidth}{!}{
  \begin{tabular}{|c|c|c|c|c|c|}
  \hline
  & Blue & Green & Magenta & Maroon & Orange \\ 
   \hline
   $\bar{\pi}_0$  &  $-0.45$ &  $-0.25$  &  $0$ &  $0.15$ &  $0.245$  \\
   \hline
   \hline
    $\Lambda_0 T_0$ & $2.456$ & $1.176$ & $1.0$ & $1.133$ & $6.017$ \\
   \hline
    $\gamma_0/\Lambda_0$ & $0.845$ & $0.475$ & $0$ & $-0.56$ & $-1.818$ \\
   \hline
  \end{tabular}
  }
  \caption{Association of initial conditions of $\bar\pi \equiv \pi/(4P)$ with the Maximum-Entropy Lagrange parameters for conformal dynamics.}
  \label{table:IC_conformal_ME}
 \end{center}
 \vspace*{-0.6 cm}
\end{table}

Figure~\ref{fig:pibar_conformal} shows the evolution of the shear inverse Reynolds number $\bar\pi = \pi/(e+P)$ as a function of the inverse Knudsen number $\tau/\tau_R$, computed from the RTA Boltzmann equation (dotted curves) and ME hydrodynamics (dashed curves), for identical initial conditions. We note excellent agreement between the microscopic kinetic and macroscopic hydrodynamic descriptions, except for the blue curves which correspond to the largest initial momentum-space anisotropy, $\bar\pi_0 = - 0.45$, where some small differences between the exact kinetic evolution and its ME hydrodynamic approximation are visible. These differences are ironed out as the inverse Knudsen number reaches values of ${\cal O}(1)$. At late times the system is close to local equilibrium; by $\tau/\tau_R \approx 3$, all curves are seen to merge with the first-order (Navier-Stokes) hydrodynamic result  $\bar{\pi}_\mathrm{NS} = \frac{4}{15} (\tau_R/\tau)$ that controls the late-time asymptotics for conformal Bjorken flow. 

\begin{figure}[!b]
\includegraphics[width=\linewidth]{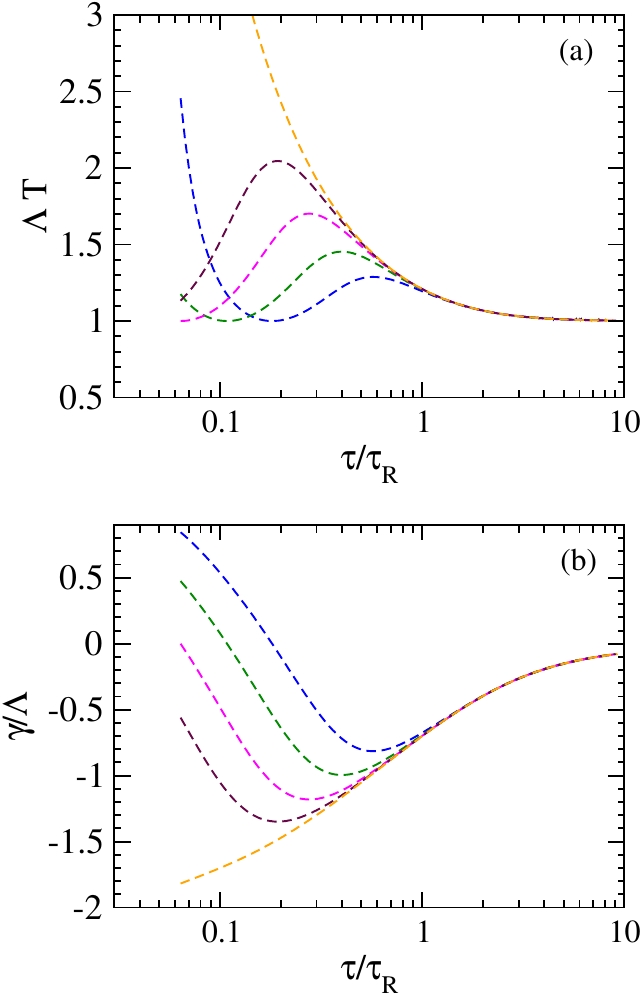}
\caption{
    Conformal ME hydrodynamic evolution of the Lagrange parameters $\Lambda$ (panel (a)) and $\gamma/\Lambda$ (panel (b)), for the maximum-entropy distribution with initial conditions listed in Table~\ref{table:IC_conformal_ME}. 
    \vspace*{-3mm}}
\label{fig:LP_conformal}
\end{figure}
    
In Figs.~\ref{fig:LP_conformal}a,b we show the ME hydrodynamic time evolution of the Lagrange parameter $\Lambda$ in units of the instantaneous inverse temperature (a), and the scaled anisotropy parameter $\bar\gamma=\gamma/\Lambda$ (b). Although different curves for $\Lambda T$ in panel (a) evolve rather differently from each other at times $\tau \ll \tau_R$ where the system is far from equilibrium, they converge to a universal curve around $\tau = \tau_R$, which then approaches unity as the system locally thermalizes ($\tau \gg \tau_R$) with $\Lambda$ assuming the role of an inverse temperature. To understand the time evolution of $\bar\gamma$ in panel (b) we first re-write the maximum-entropy distribution (\ref{fME_LRF}) with $\lambda_\Pi{\,=\,}0$ in the form
\begin{align}
    \!\!
    \fME(p, \theta_p) = \exp\Bigl[ - \Lambda \, p \, \Bigl( 1{+}\frac{\bar\gamma}{2} \Bigr)  \Bigl( 1 - \frac{3\bar\gamma}{2{+}\bar\gamma} \cos^2\theta_p \Bigr) \Bigr].
\end{align}
Here $p = \sqrt{p_T^2 + p_z^2}$ is the magnitude of the 3-momentum in the LRF and $\theta_p = \tan^{-1}(p_z/p)$. First, Eq.~(\ref{bound_Bjorken}) implies $\bar\gamma \in (-2, 1)$. For $\bar\gamma \to -2$, $\fME \to \exp(- 3 \, \Lambda \, p_z^2/p)$ whereas for $\bar\gamma \to 1$, $\fME\to\exp\bigl(- 3 \, \Lambda \, p_T^2/(2 p)\bigr)$. Thus, as $\bar\gamma$ varies between these limits, the momentum space distribution changes from one falling off steeply in the longitudinal direction ($P_L/P_T \ll 1$) to one that rapidly decreases in the transverse direction ($P_L/P_T \gg 1$). In Fig.~\ref{fig:LP_conformal}b we see that for all but the orange curve $\bar\gamma$ initially decreases rapidly towards negative values.\footnote{%
    In fact, for a free-streaming gas undergoing Bjorken expansion ($\tau_R \to \infty$), $\bar\gamma$ for these curves would approach its limit of $-2$ at late times.
}    
This is because of the initially very large longitudinal expansion rate in Bjorken flow which rapidly red-shifts the longitudinal particle momenta to small values. As a result, the effective longitudinal pressure in the fluid quickly becomes much smaller than the transverse one. The orange curve corresponds to an initial distribution where only a few particles have appreciable longitudinal momenta. Accordingly, the role of red-shifting $p_z$ by longitudinal expansion becomes negligible. Instead, microscopic collisions begin to locally isotropize the longitudinal and transverse momenta, bringing the longitudinal and transverse pressures closer to each other. The same phenomenon is observed for the other curves at somewhat later times. To describe this process of local isotropization $\bar\gamma$ increases as time proceeds, for the orange curve right away, for the others a bit later. At $\tau/\tau_R\to\infty$ the fluid reaches local thermal equilibrium (i.e. local momentum isotropy), and $\bar\gamma \to 0$.
    
\subsection{Non-conformal dynamics} 
\label{non-conformal dynamics}

We now break conformal symmetry by introducing a non-zero, fixed mass $m = 500$\,MeV for the particle constituents. For ease of comparison we keep the same conformal ansatz (\ref{conf_tau}) for the relaxation time, with $C = 10/4\pi$, as before. We then solve the RTA Boltzmann equation (\ref{RTA}) with initial conditions parametrized by the generalized Romatschke-Strickland (RS) ansatz (\ref{f_in}). $T^{\mu\nu}$ has now the three macroscopic degrees of freedom $(e, P_L, P_T)$ or, equivalently, $(T, \Pi, \pi)$. To explore the range of evolution trajectories we select a variety of RS parameter sets $(\alpha_0, T^{RS}_0, \xi_0)$, listed in Table~\ref{table_ic_mVAH}, subject to the constraint of fixed initial temperature $T_0 = 500$\,MeV and the requirement that the corresponding initial bulk and shear viscous stresses (listed in Table~\ref{table:IC}) remain inside the domain that can be accessed with $\fME$ for Fermi-Dirac statistics, as discussed in Sec.~\ref{sec:ic}.

\begin{table}[h!]
 \begin{center}
  \begin{tabular}{|c|c|c|c|c|c|c|c|}
   \hline
  &  Blue &  Green  &  Magenta &  Maroon &  Orange &  Black &  Cyan \\
   \hline
    $(\Pi/P)_0$ & 0 & $0.05$ & $-0.1$ & 0 & 0 & $0.11$ & $-0.2$ \\
   \hline
    $(\pi/P)_0$ & $-1$ & $-1$ & $-1$ & 0.99 & $-1.8$ & 0 & 0 \\
   \hline
  \end{tabular}
  \caption{Association of initial conditions $(\Pi/P)_0$, $(\pi/P)_0$ for the scaled bulk and shear viscous stresses with the colors of the curves showing their evolution in the figures below.}
  \label{table:IC}
 \end{center}
 \vspace*{-.6cm}
\end{table}
%

%
\begin{table}[h!]
 \begin{center}
	\resizebox{1.0\columnwidth}{!}{
	\begin{tabular}{|c|c|c|c|c|c|c|c|}
   		\hline
   		&  Blue &  Green  &  Magenta &  Maroon &  Orange &  Black &  Cyan \\
   		\hline
        $T^{RS}_0/T_0$ & $0.631$ & $0.921$ & $0.364$ & $1.343$ & $0.198$ & $3.598$ & $0.468$  \\
    	\hline
    	$\alpha_0$ & $0.762$ & $-0.845$ & $3.310$ & $1.726$ & $5.313$ & $-5.206$ & $3.357$ \\
    	\hline
    	$\xi_0$ &  $- 0.82$ & $- 0 .807$ & $- 0.849$ & $236.25$ & $-0.982$ & $0$ & $0$ \\
    	\hline
  	\end{tabular}
  	}
  \caption{Values of the RS parameters $(T^{RS}_0, \alpha_0, \xi_0)$ in Eq.~(\ref{f_in}) which generate the initial conditions listed in Table~\ref{table:IC}, all having the same fixed initial temperature $T_0 = 500$\,MeV and particle mass $m/T_0=1$.}
  \label{table_ic_mVAH}
 \end{center}
 \vspace*{-.6cm}
\end{table}

We start from the exact solution of the RTA Boltzmann equation, obtaining the exact temperature evolution $T(\tau)$ from the energy density given in Eq.~(\ref{T_sol_RTA}) by Landau matching, $e(\tau) = e_\mathrm{eq}\bigl(T(\tau)\bigr)$, where $e_\mathrm{eq}(T,m)$ is given by Eq.~(\ref{e_eq}). Plugging $T(\tau)$ into Eqs.~(\ref{PT_sol_RTA}, \ref{PL_sol_RTA}) we obtaine $P_T$ and $P_L$ as functions of $\tau$. From these we then calculate the exact evolution of the bulk and shear viscous stresses, $\Pi = (P_T + 2 \,P_L - 3 P)/3$ and $\pi = 2 (P_T - P_L)/3$. 

\begin{figure}[!t]
\includegraphics[width=0.95\linewidth]{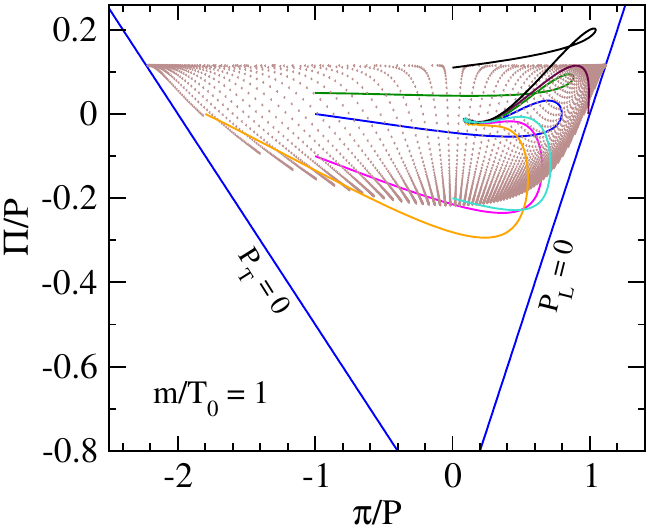}
\caption{
    Evolution of dissipative fluxes in the scaled shear-bulk plane using RTA Boltzmann equation (solid lines). The brown zone denotes the region that can be populated initially by $\fME$ for Fermi-Dirac statistics.
    \vspace*{-3mm}} 
    \label{fig:shear_bulk_evol_triangle}
\end{figure}

\begin{figure}[!b]
\includegraphics[width=0.95\linewidth]{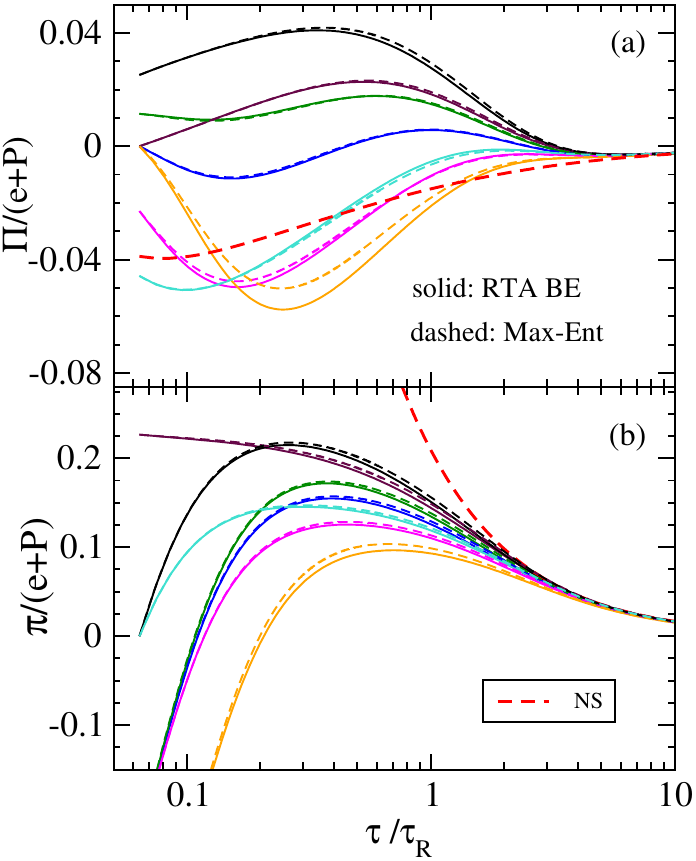}
\caption{
    Comparison of the time evolution of the (a) bulk and (b) shear inverse Reynolds numbers obtained from the exact solution of the RTA BE (solid lines) and from ME hydrodynamics (dashed lines).
    \vspace*{-3mm}} 
    \label{fig:pibar_Pibar_evol_NC}
\end{figure}

The resulting evolution trajectories are shown in Fig.~\ref{fig:shear_bulk_evol_triangle} in the $(\pi,\Pi)$-plane and by solid lines in Fig.~\ref{fig:pibar_Pibar_evol_NC} as function of time. In Fig.~\ref{fig:shear_bulk_evol_triangle} the brown dots show the range of initial conditions accessible with the ME distribution function for particles with mass $m/T_0=1$ (where here $T_0=500$\,MeV) obeying Fermi-Dirac statistics. By construction, all trajectories start inside the brown-dotted region but, since the temperature decreases with time and therefore $m/T$ increases, the ME-accessible region grows with time, and the expansion trajectories are seen to make use of this enhanced freedom. However, they never move outside the kinetically allowed\footnote{%
    Assuming positive definite distribution functions which applies for both the generalized RS and the ME parametrizations.
}
triangle delineated by the solid blue lines for zero longitudinal and transverse pressures and the condition $T^\mu_\mu \geq 0$.\footnote{%
    Positivity of the trace implies $\Pi/P \leq e/(3P) - 1$ which defines a horizontal line that moves upward as the system cools \cite{Chattopadhyay:2021ive} and is therefore not shown in the figure. We have checked that none of the trajectories ever moves above this time-evolving bound.
}
At late times all trajectories converge on the thermal equilibrium point $\bar\pi=\bar\Pi=0$.

In contrast, second-order Chapman-Enskog type hydrodynamics is known \cite{Jaiswal:2021uvv} to violate these triangular bounds for certain far-from-equilibrium initial conditions. For large values of the bulk and/or shear stresses, the off-equilibrium correction $\delta f$ (see Eq.~(\ref{deltaf_CE})) can become so large and negative that the total distribution function $f = \feq + \delta f$ becomes negative (i.e. unphysical) over a large range of momenta. This is at least a contributing factor to the observed violations. 

We will now show that the same does not happen in ME hydrodynamics. In Fig.~\ref{fig:pibar_Pibar_evol_NC} we compare the exact solution of the RTA Boltzmann equation (solid lines) with its ME hydrodynamic approximation (dashed lines), for identical initial conditions. The corresponding initial values of the Lagrange parameters $(\Lambda, \lambda_\Pi, \gamma)$ are listed in Table~\ref{table_ic_Lagrange}.
%
\begin{table}[h!]
 \begin{center}
	\resizebox{1.0\columnwidth}{!}{
	\begin{tabular}{|c|c|c|c|c|c|c|c|}
   		\hline
   		&  Blue &  Green  &  Magenta &  Maroon &  Orange &  Black &  Cyan \\
   		\hline
        $\Lambda_0 T_0$ & $0.541$ & $2.577$ & $-1.297$ & $-1.605$ & $-1.23$ & $20.182$ & $-1.835$  \\
    	\hline
    	$(\lambda_\Pi)_0 T_0$ & $0.676$ & $-1.551$ & $2.901$ & $17.399$ & $4.536$ & $-19.637$ & $3.224$ \\
    	\hline
    	$\frac{\gamma_0}
     {\Lambda_0 + \lambda_{\Pi,0}}$ &  $0.476$ & $0.421$ & $0.579$ & $-1.929$ & $0.879$ & $0$ & $0$ \\
    	\hline
  	\end{tabular}
  	}
  \caption{Initial values $\Lambda_0, (\lambda_\Pi)_0, \gamma_0$ used in ME-hydrodynamics that generate the initial conditions listed in Table~\ref{table:IC} for $T_0=500$\,MeV.}
  \label{table_ic_Lagrange}
 \end{center}
 \vspace*{-.6cm}
\end{table}

Figure~\ref{fig:pibar_Pibar_evol_NC} shows the evolution of the bulk (panel (a)) and shear (panel (b)) inverse Reynolds numbers, $\bar\Pi = \Pi/(e+P)$ and $\bar\pi = \pi/(e+P)$, as functions of the proper time in units of $\tau_R$ (i.e. of the inverse Knudsen number). (Please use Table~\ref{table:IC} to identify the initial conditions corresponding to each color.) Solid lines mark the exact RTA BE evolution, dashed lines the ME hydrodynamic approximation. The somewhat thicker red dashed curves show the evolution according to first-order Navier-Stokes hydrodynamics: $\bar{\Pi}_\mathrm{NS} = - (\zeta/s)/ (\tau T)$, $\bar{\pi}_\mathrm{NS} = (4/3) (\eta/s) / (\tau T)$. Since all curves start from the same initial temperature $T_0 = 500$\,MeV, the asymptotic Navier-Stokes result in panels (a) and (b) is unique and the same for all trajectories. In almost all cases, ranging from curves with large initial momentum {\it anisotropies} to those with large initial {\it isotropic} off-equilibrium deformations, the ME hydrodynamic results are seen to be in very good agreement with the exact RTA Boltzmann solution, throughout their evolution.\footnote{%
    The one exception are the orange trajectories, corresponding to the largest negative initial shear stress $\bar\pi_0=-1.8$ where significant deviations from the exact evolution are observed for the bulk stress. This indicates a weakness of ME hydrodynamics in handling large shear-bulk coupling effects. 
}
This is in sharp contrast with the much poorer performance of second-order Chapman-Enskog hydrodynamics, which was studied in \cite{Jaiswal:2021uvv} (see Figs.~15-17 in \cite{Jaiswal:2021uvv}) and already mentioned above.  

\begin{figure}[!b]
\includegraphics[width=0.90\linewidth]{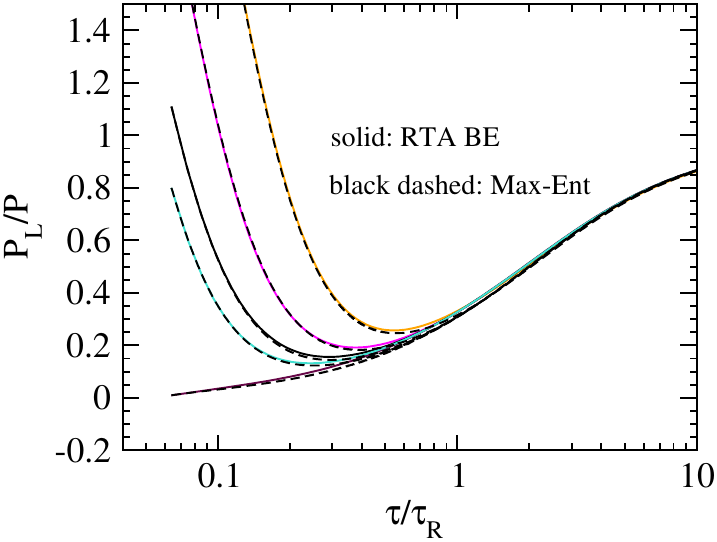}
\caption{
    Comparison of the evolution of the scaled effective longitudinal pressure obtained from kinetic theory (solid lines) and ME hydrodynamics (dashed lines).
    \vspace*{-3mm}}
    \label{fig:PL_P_evol}
\end{figure}

As explained in Sec.~\ref{conformal dynamics} during our analysis of conformal dynamics, the rapid longitudinal expansion of Bjorken expansion at early times strongly red-shifts the longitudinal momenta of particles. In the absence of isotropizing collisions among the microscopic constituents this eventually results in a distribution function that is sharply peaked in $p_z$, $f(\tau; p_T, p_z) \propto \delta(p_z)$, and the effective longitudinal pressure $P_L$ correspondingly approaches zero. Thus, generic initial conditions at early times (while the Knudsen number is large in Bjorken flow) rapidly evolve towards $P_L \approx 0$, giving rise to early-time universality in scaled quantities such as $P_L/P$. This feature being a characteristic of Bjorken expansion geometry generalises to non-conformal systems as well \cite{Chattopadhyay:2021ive, Jaiswal:2021uvv}. We show in Fig.~\ref{fig:PL_P_evol} the evolution of scaled effective longitudinal pressure as a function of scaled time, comparing the exact solution from the RTA Boltzmann equation with ME hydrodynamics. The agreement between both approaches is excellent. Fig.~\ref{fig:PL_P_evol} should be compared with panel (b) of Fig.~\ref{fig:LP_conformal} in Sec.~\ref{conformal dynamics} for its similarities; the explanation offered there directly carries over to the $P_L/P$ ratio plotted here and therefore needs no repetition.

The discussion of the ME hydrodynamic time evolution of the energy-momentum tensor is completed in Appendix~\ref{appc} with an analysis of the evolution of the ME Lagrange parameters that accompanies the evolution trajectories shown in Figs.~\ref{fig:pibar_Pibar_evol_NC} and \ref{fig:PL_P_evol}. We close this section by comparing the performance of ME hydrodynamics with that of the (modified) viscous anisotropic hydrodynamic (\mVAH) approach discussed in Ref.~\cite{Jaiswal:2021uvv}. Like ME hydrodynamics, \mVAH{} was found to agree very well with the exact RTA BE solutions shown in Figs.~\ref{fig:pibar_Pibar_evol_NC} and \ref{fig:PL_P_evol} (see Figs.~18-20 in \cite{Jaiswal:2021uvv}). In fact, in Appendix~\ref{appa} we show that for generalized RS initial conditions, as used in Figs.~\ref{fig:pibar_Pibar_evol_NC} and \ref{fig:PL_P_evol}, the agreement of \mVAH{} with the exact RTA BE solution is even better than that of ME hydrodynamics. On the other hand, if we initialize the identical macroscopic initial condition listed in Table~\ref{table:IC} not with a generalized RS ansatz for the distribution function (which is used to close the \mVAH{} equations), with parameters listed in Table~\ref{table_ic_mVAH}, but instead with a maximum entropy distribution $\fME$ with parameters listed in Table~\ref{table_ic_Lagrange} (which is used to close the ME hydrodynamic equations), the picture turns upside-down: For $\fME$ initial conditions, the RTA BE evolution is slightly different than for \mVAH{} initial conditions, and ME hydrodynamics describes the exact evolution more accurately than \mVAH. 

In spite of the excellent ME hydrodynamic description of the exact evolution of the energy-momentum tensor obtained from kinetic theory, it is shown in App.~\ref{appd} that significant discrepancies between the micro- and macroscopic approaches are seen in the evolution of the entropy density. This parallels a similar finding for anisotropic hydrodynamics, first made in \cite{Chattopadhyay:2018apf} and here confirmed in 
App.~\ref{appd}. In fact, for the entropy evolution in Bjorken flow, ME hydrodynamics and \mVAH{} agree much better with each other than either does with the exact kinetic theory.

In summary, for Bjorken expansion \mVAH{} and ME hydrodynamics are both highly competitive macroscopic approximations to the underlying kinetic evolution. However, since the modified RS ansatz (\ref{f_in}) for the distribution function (on which \mVAH{} rests) was custom-built for Bjorken geometry, whereas this is not the case for the $\fME$ ansatz (\ref{MaxEnt distribution}), ME hydrodynamics is expected to exhibit superior performance in general expansion scenarios, without the restricting symmetries of Bjorken flow. 

\vspace*{-2mm}
\section{Gubser flow}
\label{sec3}
\vspace*{-3mm}

To put this expectation to the test, in this section we make a first step in this direction by studying Gubser flow \cite{Gubser:2010ze, Gubser:2010ui}. While Bjorken flow, without any transverse expansion, is widely assumed to be a good approximation for the dynamical state of the matter formed in ultra-relativistic heavy-ion collisions just after its creation, the finite transverse size of the colliding nuclei implies large transverse density and pressure gradients of the created matter which, after a period of a few relaxation times, drive  collective transverse expansion, starting at the edges of the transverse energy density distribution. The subsequent stage of fully three-dimensional expansion without any remaining symmetries and very different longitudinal and transverse expansion rates can no longer be treated analytically. Gubser flow is an idealization located somewhere between Bjorken flow and generic three-dimensional flow, by incorporating transverse flow on top of longitudinal boost-invariant expansion, albeit with a very specific transverse flow profile\footnote{%
    The transverse expansion encoded in Gubser flow is so violent that, at late times, the  {\it transverse} momenta of the constituent particles are red-shifted all the way towards zero effective transverse pressure $P_T$. 
} 
that retains just enough symmetry that the RTA Boltzmann equation continues to be exactly solvable by analytic means \cite{Denicol:2014xca, Denicol:2014tha}.

Gubser derived his flow profile by starting from Bjorken flow, keeping longitudinal boost-invariance and reflection symmetry as well as azimuthal rotational symmetry around the beam axis but relaxing the assumption of transverse homogeneity. Mathematically speaking, the Gubser flow profile replaces the $ISO(2) \otimes SO(1,1) \otimes Z_2$ symmetry of Bjorken flow by invariance under $SO(3)_q \otimes SO(1,1) \otimes Z_2$, where $SO(3)_q$ denotes the special conformal group of transformations \cite{Gubser:2010ze, Gubser:2010ui}. The symmetries of this flow are manifest in de Sitter coordinates in a curved space-time constructed as the direct product of 3-dimensional de Sitter space with a line, $dS_3 \otimes R$. One first Weyl-rescales the Milne metric,
\begin{equation}
\label{Weyl}
    d\hat{s}^2 \equiv \frac{ds^2}{\tau^2} = \frac{d\tau^2 - dr^2 - r^2 d\phi^2}{\tau^2} - d\eta^2,
\end{equation}
and then transforms the Milne coordinates $x^\mu = (\tau, r, \phi, \eta)$ to ``Gubser coordinates" $\hat{x}^\mu = (\rho, \theta, \phi, \eta)$, with
\begin{align}
    \rho &= - \sinh^{-1} \left( \frac{1 - q^2 \tau^2 +q^2 r^2}{2 q \tau} \right), 
\label{Gubser-rho}\\\label{Gubser-theta}
    \theta &= \tan^{-1} \left( \frac{2qr}{1 + q^2 \tau^2 - q^2 r^2} \right),
\end{align}
where $q$ is an energy scale that sets the transverse size of the system. In these coordinates the metric takes the form
\begin{equation}
    \hat{g}_{\mu\nu} = \mathrm{diag} \left( 1, \, - \cosh^2\rho, \, - \cosh^2\rho \, \sin^2\theta, \, - 1 \right)
\end{equation}
with the line element
\begin{align}
    d\hat{s}^2 = d\rho^2 - \cosh^2\rho \left(d\theta^2 + \sin^2\theta \, d\phi^2 \right) - d\eta^2,
\end{align}
which is manifestly invariant under the $SO(3)_q$ group of rotations in $(\theta, \phi)$ space. In these coordinates, the flow appears static, $\hat{u}^\mu = (1, 0,0,0)$, and all quantities depend only on the Gubser time $\rho$. Moreover, to make conformal symmetry manifest, all quantities expressed in Gubser coordinates (denoted by a hat) are rendered dimensionless by appropriate rescaling with powers of the Weyl rescaling parameter $\tau$. For example, the Gubser temperature and energy-momentum tensor are
\begin{equation}
    \!\!
    \hat{T}(\rho) = \tau \,  T(\tau,r), \quad\hat{T}^{\mu \nu}(\rho) = \tau^2 \, \frac{\partial \hat{x}^\mu}{\partial x^\alpha} \frac{\partial \hat{x}^\nu}{\partial x^\beta} \, T^{\alpha \beta}(\tau,r).
\end{equation}
The Gubser symmetries imply a diagonal form of the energy-momentum tensor, $\hat{T}^{\mu\nu}{\,=\,}\mathrm{diag}(\hat{e}, \hat{P}_T, \hat{P_T}, \hat{P}_L)$, with $\hat{e}{\,=\,}3 \hat{P}{\,\propto\,}\hat{T}^4$ and zero trace such that there is no bulk viscous pressure and $\hat{e} = 2 \hat{P}_T + \hat{P}_L$. The shear stress tensor has only one independent component which we take (as in Bjorken flow) to be $\hat{\pi} = \frac{2}{3}\,(\hat{P}_T{-}\hat{P}_L)$. Using as before bars to denote normalization by $4P$, in kinetic theory the normalized shear stress spans the range between $\hat{\bar\pi}{\,=\,}{-}\frac{1}{2}$ (where $\hat{P}_T{\,=0\,}$ and $\hat{P}_L{\,=\,}3\hat{P}$) and $\hat{\bar\pi}{\,=\,}+\frac{1}{4}$ (where $\hat{P}_L{\,=\,}0$ and $\hat{P}_T{\,=\,}\frac{3}{2}\hat{P}$). 

The phase-space distribution function satisfying the symmetries of this flow can only depend on Gubser-invariant variables, $\rho$, $\hat{p}_{\Omega} \equiv \sqrt{\hat{p}^2_\theta + \hat{p}^2_\phi/\sin^2\theta }$, 
and $\hat{p}_{\eta}$, i.e., $f(x,p) = f(\rho; \hat{p}_{\Omega}, \hat{p}_\eta)$ \cite{Denicol:2014tha}. Using these choices of phase-space variables, the RTA BE simplifies to
\begin{equation}{\label{RTA_Gubser}}
    \frac{\partial f}{\partial \rho} = - \frac{1}{\hat{\tau}_R} \, \left( f - \feq \right)
\end{equation}
where the relaxation time $\hat{\tau}_R = 5 (\eta/s) / \hat{T}$. The equilibrium distribution is
\begin{equation}
    \feq = \exp\left(- \frac{\hat{p}^\rho}{\hat{T}} \right) \quad\text{with}\quad 
    \hat{p}^\rho \equiv \sqrt{\frac{\hat{p}^2_\Omega}{\cosh^2\rho} + \hat{p}^2_\eta}.
\end{equation}
The formal solution of Eq.~(\ref{RTA_Gubser}) is \cite{Denicol:2014xca}
\begin{align}
    \!\!\!\!
     & f(\rho; \hat{p}^2_\Omega, \hat{p}_\eta) = D(\rho, \rho_0) \, f_0(\hat{p}^2_\Omega, \hat{p}_\eta) \nonumber \\
     & \qquad \qquad
     + \int_{\rho_0}^{\rho} \, \frac{d\rho'}{\hat{\tau}_R(\rho')} \, D(\rho, \rho') \, \feq\left(\hat{p}^\rho(\rho'), \hat{T}(\rho) \right),
\end{align}
with $D(\rho_2, \rho_1) = \exp\bigl(- \int_{\rho_1}^{\rho_2} \, d\rho'/\hat{\tau}_R(\rho')\bigr)$ for the damping function. Similar to the Bjorken case we parametrize the initial distribution with a Romatschke-Strickland ansatz
\begin{align}
    \label{f0_Gubser}
    \!\!\!\!
    f_0(\hat{p}^2_\Omega, \hat{p}_\eta) = \exp\left(- \frac{\sqrt{\hat{p}^2_\Omega/\cosh^2\rho_0 + \left(1{+}\xi_0 \right) \hat{p}^2_\eta}}{\hat{T}^{RS}_0} \right).
\end{align}
The temperature evolution is then obtained by solving the integral equation\footnote{%
    Equation~(\ref{T_hat_sol}) extends the original work in \cite{Denicol:2014xca, Denicol:2014tha} from equilibrium initial conditions to the general case of non-zero initial momentum anisotropy $\xi_0$ \cite{Nopoush:2014qba, Martinez:2017ibh}.
}
\begin{align}\label{T_hat_sol}
    \hat{T}^4(\rho) &= D(\rho, \rho_0) \, (\hat{T}^{RS}_0)^4 \, {\cal E}_{G}(\rho, \rho_0; \xi_0) 
\nonumber \\
    &+ \int_{\rho_0}^{\rho} \, \frac{d\rho'}{\hat{\tau}_R(\rho')} \, D(\rho, \rho') \, \hat{T}^4(\rho') \, {\cal E}_{G}(\rho,\rho'; 0)
\end{align}
where
\begin{align}
    {\cal E}_{G}(\rho; \rho_1, \xi) = \left( \frac{\cosh\rho_1}{\cosh\rho} \right)^4 {\cal H}_e\left(\frac{\cosh\rho}{\cosh\rho_1 \, \sqrt{1{+}\xi}} \right),
\end{align}
with ${\cal H}_e(x)$ defined in Eq.~(\ref{H_e}). With $\hat{T}(\rho)$ from (\ref{T_hat_sol}) the exact distribution function can be computed from Eq.~(\ref{RTA_Gubser}). 

\subsection{ME hydrodynamics for Gubser flow}
\label{sec3a}

To derive the Maximum Entropy hydrodynamic equations for Gubser flow we first write down the exact evolution equations for the two independent components of $T^{\mu\nu}$, $e$ and $P_T$:\footnote{%
    This choice differs from what we did for Bjorken flow where, in the conformal limit, we selected $e$ and $P_L$. The reason is that in Bjorken flow the thermal energy decreases with time exclusively by work done by the {\it longitudinal} pressure $P_L$ whereas in Gubser flow we want to focus on the effects of the {\it transverse} pressure $P_T$ on the cooling and flow patterns of the system.
}
\begin{align}
    \frac{d\hat{e}}{d\rho} &= - 2\tanh\rho \, \left( \hat{e} + \hat{P}_T \right), \label{e_hat_evol}\\
    \frac{d\hat{P}_T}{d\rho} &= - \frac{1}{\tau_R} \left(\hat{P}_T - \hat{P} \right) - \left( 2 \tanh\rho \right) \,  \hat{\zeta}^\perp. \label{PT_hat_evol}
\end{align} 
Note that $2\,\tanh\rho$ is the scalar expansion rate in Gubser flow, $\hat\theta\equiv\hat\nabla\cdot\hat{u}=2\,\tanh\rho$, which here takes the place that $1/\tau$ holds in Bjorken flow.\footnote{%
    We refer to Appendix~\ref{appe} for a discussion of the nontrivial relationship between the scalar expansion rates $\theta=\nabla\cdot{u}$ in Minkowski space and $\hat\theta\equiv\hat\nabla\cdot\hat{u}$ in Gubser space. We note in particular that $\hat\theta=2\,\tanh\rho$ can take either sign whereas the scalar expansion rate in Minkowski space is always positive for Gubser flow, $\theta\geq0$.
}
The coupling $\hat{\zeta}^{\perp}$ is defined by 
\begin{equation}
    \label{zeta_hat}
    \hat{\zeta}^\perp = 2 \hat{P}_T - 2 \hat{I}^{\mathrm{exact}}_{202}
\end{equation}
with the thermodynamic integral 
\begin{equation}
\label{moments_Gubser}
    \hat{I}^\mathrm{exact}_{nrq} \equiv \frac{1}{(2q)!!} \int d\hat{P}\, \left(\hat{p}^\rho \right)^{n-r-2q} \,  \left(\hat{p}_\eta \right)^r  \, \left( \frac{\hat{p}_\Omega}{\cosh\rho} \right)^{2q} \, f,
\end{equation}
where 
\begin{equation}
    d\hat{P} \equiv \frac{d\hat{p}_\theta \, d\hat{p}_\phi \, d\hat{p}_\eta }{(2\pi)^3 \, \hat{p}^\rho \, \sqrt{-\hat{g}}},
\end{equation} 
with $\sqrt{-\hat{g}} = \cosh^2\rho \, \sin\theta $.
Note that Eq.~(\ref{moments_Gubser}) maps onto Eq.~(\ref{I_exact}) with the substitutions
\begin{equation}
\label{AdS_to_Cartesian}
    \frac{\hat{p}_\theta}{\cosh\rho} \mapsto p_{x}, \,\, \frac{\hat{p}_\phi}{\cosh\rho \, \sin\theta} \mapsto p_{y}, \,\, \hat{p}_\eta \mapsto p_{z},
\end{equation}
where the use of LRF coordinates is implied.\footnote{%
    Note that this mapping also implies $\hat{p}_\Omega/\cosh\rho \mapsto p_T$ and $\hat{p}^\rho \mapsto E_p = p$.
}

As in Sec.~\ref{secIIB} we close the set of exact equations (\ref{e_hat_evol}-\ref{PT_hat_evol}) by replacing $\hat{I}^{\mathrm{exact}}_{202} \to \hat{\tilde{I}}_{202}$ where the tilde indicates substitution of the Maximum Entropy distribution $\fME$ as an approximation for the exact solution $f$ of the RTA Boltzmann equation:
\begin{equation}
\label{moments_Gubser_2}
    \hat{\tilde{I}}_{nrq} \equiv \frac{1}{(2q)!!} \int d\hat{P} \, \left(\hat{p}^\rho \right)^{n-r-2q} \,  \left(\hat{p}_\eta \right)^r  \, \left( \frac{\hat{p}_\Omega}{\cosh\rho} \right)^{2q} \, \fME.
\end{equation}
In Gubser coordinates the ME distribution reads
\begin{equation}
\label{fME_Gubser}
    \fME(\rho, \hat{p}_\Omega, \hat{p}_\eta) = \exp\left[ - \hat{\Lambda} \, \hat{p}^\rho - \frac{\hat{\gamma}}{\hat{p}^\rho} \left( \frac{\hat{p}^2_\Omega}{2 \cosh^2\rho} - \hat{p}^2_\eta \right) \right].
\end{equation}
We will again solve directly for the Lagrange parameters $(\hat{\Lambda}, \hat{\gamma})$ instead of $(\hat{e},\hat{P}_T)$, by inverting
\begin{align}
    \begin{pmatrix}
         d\hat{e} \\ 
         d \hat{P}_T
    \end{pmatrix} = 
\begin{pmatrix}
     -\hat{\tilde{I}}_{300} & - \hat{\tilde{I}}_{301}{+}\hat{\tilde{I}}_{320}  \\
    - \hat{\tilde{I}}_{301} & - 2 \hat{\tilde{I}}_{302}{+}\hat{\tilde{I}}_{321} \\ 
\end{pmatrix}
\begin{pmatrix}
     d\hat{\Lambda} \\ 
     d\hat{\gamma}
\end{pmatrix}.
\end{align}
Making use of the mapping (\ref{AdS_to_Cartesian}) we compute the moments $\hat{\tilde{I}}_{nrq}$ from Eq.~(\ref{Inrq_conformal}) by replacing the Lagrange parameters $(\Lambda,\, \bar\gamma)$ in the latter by their hatted counterparts.

\subsection{Chapman-Enskog\,hydrodynamics\,for\,Gubser\,flow}
\label{sec3b}

In order to demonstrate the degree of improvement attained by using the ME truncation scheme in comparison with traditional hydrodynamic approaches, we briefly recap the evolution equations for the latter. For illustration we use the Gubser flow version of the Chapman-Enskog-like framework briefly discussed in Sec.~\ref{sec_CE_intro}. Recall that Gubser flow is conformally symmetric and thus the bulk viscous pressure vanishes, $\hat\Pi=0$. Up to third order in the Chapman-Enskog (CE) approximation, the Gubser flow evolution equations for the energy density and shear stress\footnote{%
    Note that the definitions of $\hat\pi$ here and in \cite{Chattopadhyay:2018apf} differ by a sign.
}
are \cite{Chattopadhyay:2018apf}
\begin{align}
    \frac{d\hat{e}}{d\rho} &= - 2 \tanh\rho \left(\frac{4}{3} \, \hat{e} + \frac{\hat{\pi}}{2} \right), 
\label{e_sol_CE_Gubser}
\\
    \frac{d\hat{\pi}}{d\rho} &= - \frac{\hat{\pi}}{\hat{\tau}_R} - \tanh\rho \, \left( \frac{4}{3}\hat{\beta}_\pi + \hat{\lambda} \, \hat{\pi} - \hat{\chi} \, \frac{\hat{\pi}^2}{\hat{\beta}_\pi} \right), 
\label{pi_sol_CE_Gubser}
\end{align}
with first- and second-order transport coefficients $\hat{\beta}_\pi = 4 \hat{P}/5$ and $\hat{\lambda} = 46/21$. The last term in Eq.~(\ref{pi_sol_CE_Gubser}) enters only at third order in the CE expansion, with third-order transport coefficient $\hat{\chi} = 72/245$.\footnote{%
    Note that the second-order Chapman-Enskog equation for shear evolution, i.e. Eq.~(\ref{pi_sol_CE_Gubser}) with $\hat{\chi} = 0$, is identical to the corresponding evolution in the DNMR framework \cite{Denicol:2014xca}.
}

\subsection{Gubser evolution in kinetic theory and hydrodynamics}
\label{sec3c}

We solve the RTA Boltzmann equation exactly for three different shear viscosity values, $4\pi \eta/s \in (1, 3, 10)$. We start the evolution at Gubser time $\rho_0 = -10$ and tune the Romatschke-Strickland parameters $(\hat{T}^{RS}_0, \xi_0)$ such that the initial Gubser temperature in all cases is fixed at $\hat{T}_0 = 0.002$.\footnote{%
    In the center of a fireball with typical transverse size $1/q \approx 4.3$\,fm, this corresponds to an actual temperature $T_0 \approx 2$\,GeV at Milne time $\tau_0 \approx 1.9 \times 10^{-4}$\,fm/$c$.
}
For a better visual separation of the evolution trajectories, the initial values for the normalised shear stress $\hat{\bar\pi}_0$ for $4\pi\eta/s = 1,\ 3,$ and $10$ are set to $-0.45$, $- 0.25$, and $0$, respectively. The initial ME Lagrange parameters $(\hat\Lambda_0, \hat{\gamma})$ are tuned to reproduce these initial values of $\hat{T}_0$ and $\hat{\bar\pi}_0$. All these initial conditions are summarised in Table~\ref{table:IC_conformal_aH_Gubser}.\footnote{%
    Note that the values of $\xi_0$ in Table \ref{table:IC_conformal_aH_Gubser} are identical to those in Table \ref{table:IC_conformal_aH} for conformal Bjorken flow. In both cases we generate identical initial normalised shear stresses; in conformal dynamics these depend only on the anisotropy parameter $\xi_0$, irrespective of the temperature scale ($T^{RS}_0$ or $\hat{T}^{RS}_0$) in the RS-ansatz. 
}   
%
%
\begin{table}[h!]
 \begin{center}
  \begin{tabular}{|c|c|c|c|}
   \hline
   $\hat{\bar{\pi}}_0$  &  $-0.45$ &  $-0.25$  &  $0$   \\
   \hline
    $ \hat{T}^{RS}_0 $ & $8.228 \times 10^{-4}$ & $1.437 \times 10^{-3}$ & $2 \times 10^{-3}$ \\
   \hline
    $\xi_0$ & $- 0.985$ & $-0.828$ & $0$  \\
   \hline
   \hline
   $ \hat{\Lambda}_0 $ & $1227.9$ & $587.97$ & $500$ \\
   \hline
    $\hat{\gamma}_0$ & $1037.16$ & $279.2$ & $0$  \\
   \hline
  \end{tabular}
  \caption{Romatschke-Strickland parameters $(\hat{T}^{RS}_0, \xi_0)$ and Maximum Entropy Lagrange parameters ($\hat{\Lambda}_0, \hat{\gamma}_0$) used to initialise the RTA Boltzmann and ME hydrodynamic equations, respectively, at $\hat{T}_0 = 0.002$ and the normalised shear stresses $\hat{\bar\pi}_0$ shown. The initial conditions shown in columns 2, 3, and 4 are evolved with specific shear viscosities $4\pi\eta/s=1$, 3, and 10, respectively.}
  \label{table:IC_conformal_aH_Gubser}
 \end{center}
\end{table}
%

\begin{figure}[t!]
\includegraphics[width=0.95\linewidth]{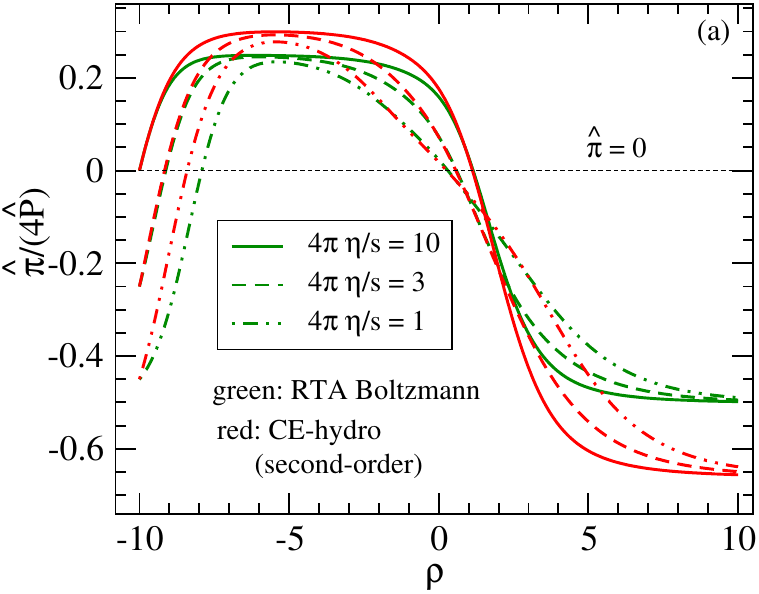}
\includegraphics[width=0.95\linewidth]{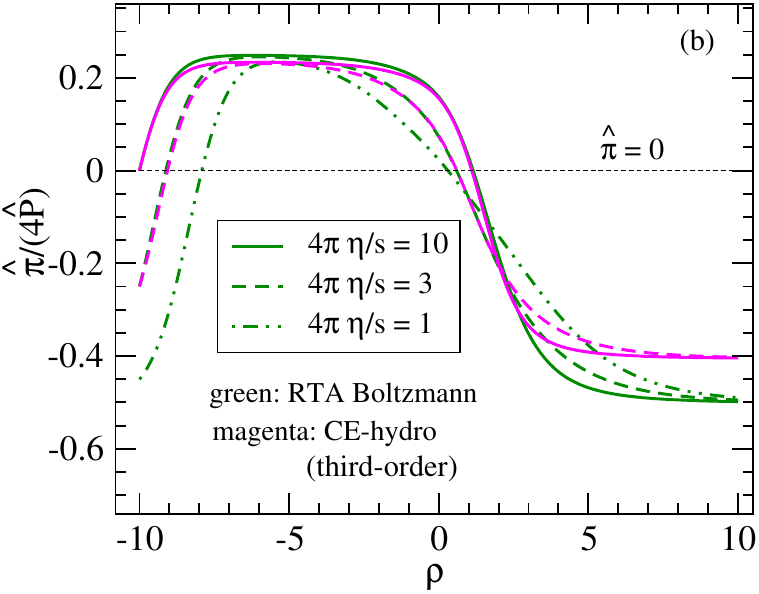}
\caption{
     Gubser evolution of the normalised shear stress, comparing the exact solution of the RTA Boltzmann equation (green lines) with CE hydrodynamics at second (red lines in panel a) and third (magenta lines in panel b) order of the Chapman-Enskog expansion. Different line styles correspond to different shear viscosities as detailed in the legend.
     } 
     \label{fig:CE_Gubser}
\end{figure} 

To set expectations we first compare in Fig.~\ref{fig:CE_Gubser} the exact RTA BE results from microscopic kinetic theory (green lines) with the macroscopic approximations of second- (red lines, panel (a)) and third-order (magenta lines, panel (b) -- see also Fig.~4 in \cite{Chattopadhyay:2018apf}) Chapman-Enskog hydrodynamics. Three different line styles distinguish evolutions with different shear viscosities as shown in the legend. 

As already discussed in \cite{Chattopadhyay:2018apf}, in kinetic theory the evolution of the normalized shear stress in Gubser flow is controlled by an ``early-time''\footnote{%
    We emphasize that Gubser ``time'' $\rho$ and Milne time $\tau$ are connected by the position-dependent relation (\ref{Gubser-rho}) and provide very different intuitions about the system's evolution \cite{Gubser:2010ui, Gubser:2010ze, Denicol:2014xca, Denicol:2014tha}.
}
attractor at large negative Gubser time $\rho$, corresponding to $\hat{\bar{\pi}}=\frac{1}{4}$ (i.e. the longitudinal free-streaming limit $\hat{P}_L=0$), and a ``late-time'' attractor at large positive $\rho$, corresponding to $\hat{\bar\pi}=-\frac{1}{2}$ (i.e. the transverse free-streaming limit $\hat{P}_T=0$). For all three initial conditions and specific shear viscosities the ``early-time'' dynamics is characterized by a rapid approach towards the {\it longitudinal} free-streaming limit at $\hat{P}_L=0$, very similar to what is observed in Milne time in Fig.~\ref{fig:PL_P_evol} for Bjorken flow. Around $\rho=0$ the Gubser dynamical evolution of the shear stress is non-universal and depends quite sensitively on the value of the specific shear viscosity $\eta/s$. The ``late-time'' behavior, however, is again universal and characterized by an approach to the {\it transverse} free-streaming limit at zero {\it transverse} pressure, $\hat{P}_T=0$. This has no analog in Bjorken flow and must therefore be caused by the transverse expansion in Gubser flow.

That the fluid dynamics for Gubser flow approaches free-streaming limits, characterized by very large Knudsen numbers Kn${\,=\,}\hat{\tau}_R|\hat{\theta}|{\,=\,}2\hat\tau_R|\tanh\rho|$, at both large negative and large positive $\rho$ values is implicit in Fig.~4 of Ref.~\cite{Denicol:2014tha} which shows the Knudsen number growing exponentially in both limits. What was not realized in that first analysis is that at large negative $\rho$ the scalar expansion rate $|\hat\nabla\cdot\hat{u}|$ is dominated by {\it longitudinal} expansion whereas the growth of the Knudsen number at large positive $\rho$ has to be associated with a large (relative to the microscopic scattering rate) {\it transverse} expansion rate.\footnote{%
    We refer to App.~\ref{appe} for technical details.
}
This explains the approach to different attractors ($\hat{P}_L=0$ at negative $\rho$, $\hat{P}_T=0$ at positive $\rho$) at early and late Gubser times.

The red and magenta lines in Figs.~\ref{fig:CE_Gubser}a and b show that CE hydrodynamics does not correctly reproduce either one of these two attractors. The discrepancy between the exact kinetic theory and its macroscopic hydrodynamic approximation gets smaller at third order of the CE expansion (panel (b)) than at second order (panel (a)),\footnote{%
    Note that $\hat{\bar\pi}{\,>\,}0.25$ ($\hat{\bar\pi}{\,<\,}{-}0.5$) implies $\hat{P}_L{\,<\,}0$ ($\hat{P}_T{\,<\,}0$). At second order in the CE expansion, CE hydrodynamics for Gubser flow thus violates basic kinetic theory limits. 
}
but clearly remains sizeable \cite{Chattopadhyay:2018apf}.\footnote{%
    Notwithstanding this improvement, the third-order theory breaks down for large negative initial shear. This is why in Fig.~\ref{fig:CE_Gubser}b there is no magenta curve for $\hat{\bar\pi}_0 = - 0.45$.
} 
The transition around $\rho=0$ between the ``early" and ``late" free-streaming limits, where the deviations from thermal equilibrium are small, is described well by CE hydrodynamics, at both second and third order precision. The duration of agreement increases with the strength of the microscopic interactions (smaller $\eta/s$).

\begin{figure}[t]
\includegraphics[width=0.8\linewidth]{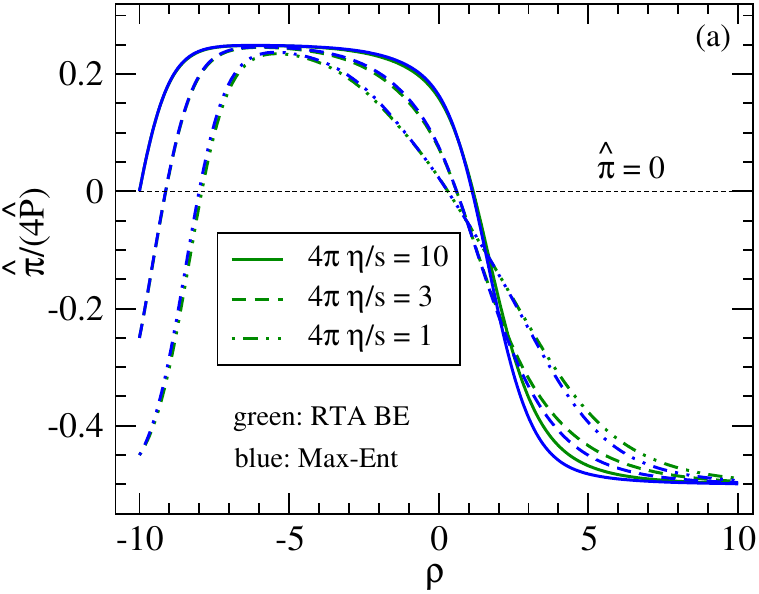}
\includegraphics[width=0.8\linewidth]{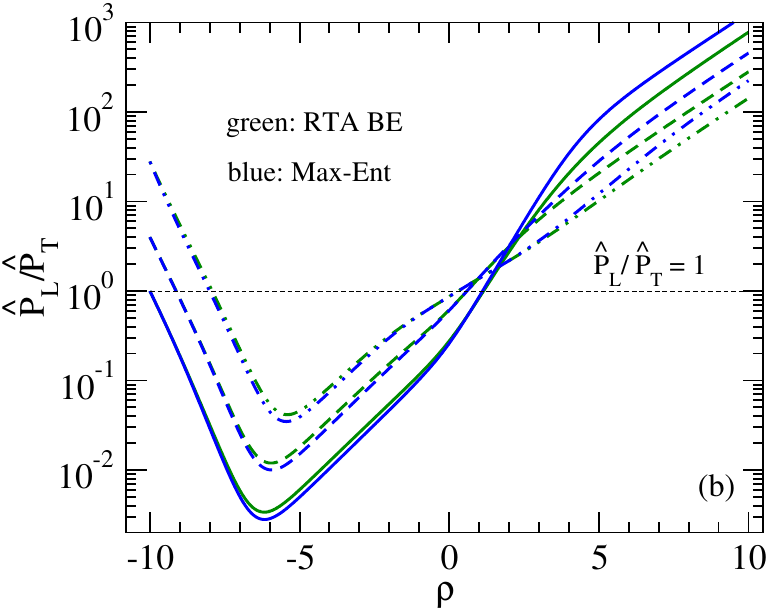}
\caption{
    Evolution of (a) the shear inverse Reynolds number and (b) the pressure anisotropy in Gubser flow, comparing the exact solution of the RTA Boltzmann equation (green curves) with ME hydrodynamics (blue curves). Where the green curves become invisible they are hidden behind the blue curves.
    } 
    \label{fig:ME_hydro_Gubser}
\end{figure}

Figure~\ref{fig:ME_hydro_Gubser} shows that the shortcomings CE hydrodynamics are not shared by Maximum Entropy hydrodynamics. For $\rho{\,<\,}0$ the evolution of the normalised shear stress shown in panel (a) agrees almost perfectly between the microscopic and macroscopic approaches. At positive $\rho$ small differences can be seen (the exact shear stress lies slightly above the ME approximation) but at $\rho\to\infty$ the normalized shear stress from ME hydrodynamics converges perfectly to the correct asymptotic value for transverse free-streaming. The kinetic constraints $\hat{P}_{L,T}{\,\geq\,}0$ are never violated. Studying the pressure anisotropy $\hat{P}_L/\hat{P}_T$ in panel (b) reveals that, as $\hat{e}{\,=\,}3\hat{P}\to0$ at large $\rho$, the longitudinal-to-transverse pressure ratio $\hat{P}_L/\hat{P}_T$ grows exponentially but differs by a constant factor between kinetic theory and ME hydrodynamics; this constant approaches 1 (i.e. the difference between the green and blue curves vanishes) as the specific shear viscosity decreases, i.e. as the system becomes more strongly coupled.\footnote{%
    As already remarked at the end of Sec.~\ref{sec_mehydro_intro}, the curves in Fig.~\ref{fig:ME_hydro_Gubser} agree with those first shown in Ref.~\cite{Calzetta:2019dfr}, claimed (incorrectly) to result from a hydrodynamic framework based on the principle of maximizing the {\it rate  of entropy production}. The correct interpretation of the numerical results in \cite{Calzetta:2019dfr} is that they are predictions of ME hydrodynamics, i.e. they maximize the {\it entropy} itself. In fact, almost everywhere in  Fig.~\ref{fig:ME_hydro_Gubser} the deviations from local thermal equilibrium are so large that the approximations made in \cite{Calzetta:2019dfr} to arrive at their final form for $f_\mathrm{DTT}$ fail catastrophically.
    }  

In Fig.~4 of Ref.~\cite{Chattopadhyay:2018apf} a version of Fig.~\ref{fig:ME_hydro_Gubser} was shown where all blue ME hydrodynamic curves were replaced by trajectories obtained from anisotropic hydrodynamics (\texttt{aHydro}). We have repeated that exercise for \mVAH{} (not shown) and found excellent agreement between the \mVAH{} and ME-hydrodynamic predictions. As we had noticed in Sec.~\ref{non-conformal dynamics} and discuss in detail in App.~\ref{appa} for Bjorken flow, for the RS-type initial conditions used in Fig.~\ref{fig:ME_hydro_Gubser} the \mVAH{} predictions for Gubser flow again agree slightly better with the exact kinetic evolution than the ones from ME hydrodynamics. (This may flip again for ME initial conditions but we did not solve the RTA Boltzmann equation for that case.) So ME hydrodynamics and \mVAH{} are again very competitive hydrodynamic approximations of the underlying kinetic theory when considering Gubser flow.

So why did the anisotropic hydrodynamic approach (whose construction made heavy use of the specific symmetries of Bjorken flow) not fail ---as one might have expected--- when moving from Bjorken flow (without any transverse expansion) to Gubser flow (with very strong transverse expansion that completely dominates the fluid dynamics at late Gubser times)? The answer is that \mVAH{} for Gubser flow is not the same as \mVAH{} for Bjorken flow. Although both are obtained by using an (almost identical looking) RS-type ansatz for the microscopic distribution function in order to close the hydrodynamic equations, for Gubser flow the RS distribution function (\ref{f0_Gubser}) is expressed through Gubser coordinates instead of the Milne coordinates used in (\ref{f_in}), which span a different type of space-time: one is intrinsically curved, the other flat. This makes them physically very different distributions. In the end the ``Gubser RS ansatz" (\ref{f0_Gubser}) is as well adapted to Gubser flow as the standard RS ansatz (\ref{f_in}) is to Bjorken flow, sharing this feature with the ME ansatz (Eqs.~(\ref{f_ME_Bjorken}) and (\ref{fME_Gubser}), respectively).

\section{Conclusions and outlook}
\label{sec_conclusions}

Using the Maximum Entropy distribution constrained by the full energy-momentum tensor to truncate the moment hierarchy of the relativistic Boltzmann equation in Relaxation Time Approximation, we here developed {\it ME hydrodynamics}, a new relativistic framework for dissipative fluid dynamics that accounts non-perturbatively for the full set of dissipative energy-momentum flows. The framework can be straightforwardly extended to systems with conserved charges, by including the corresponding diffusion currents as additional constraints when maximizing the Shannon entropy, but in this work we focused on fluids without such conserved charges. ME hydrodynamics is conceived to provide an extension of standard second-order (``transient'') relativistic dissipative fluid dynamics, which is based on the assumption of weak dissipative flows, into the domain of far-from-equilibrium dynamics. It is a generically macroscopic approach which uses only macroscopic hydrodynamic information, in the sense that the distribution function used for moment truncation is only constrained by macroscopic hydrodynamic quantities. It makes use of microscopic Boltzmann kinetic theory only to the extent that it is assumed that {\it some} such kinetic description exists for the fluid, without additional specifics. Furthermore, the kinetic description is only used to determine the coupling to non-hydrodynamic
moments; the form of the evolution equations for the conserved charges remains completely general.

While the framework accommodates arbitrary three-dimensional flow patterns, we here studied it, for testing purposes, only for Bjorken and Gubser flow. Assuming that the microscopic physics of the fluid is controlled by the RTA Boltzmann equation, these flows provide highly symmetric environments in which this underlying microscopic physics can be solved semi-analytically with arbitrary numerical precision. These microscopic solutions then provide the exact space-time evolution of the full energy-momentum tensor of the system against which the predictions obtained numerically from the macroscopic ME hydrodynamic framework can be compared with quantitative precision. In this work we performed such comparisons for both massless and massive Boltzmann gases undergoing Bjorken expansion and for a massless gas undergoing Gubser flow. The agreement of the macroscopic ME hydrodynamic predictions with the exact underlying kinetic results was found to be excellent in all cases, except for initial conditions encoding the most extreme deviations from local thermal equilibrium where differences of a few percent were visible between the micro- and macroscopic descriptions.

As shown in this work and in earlier publications, the same is not true for most other macroscopic hydrodynamic theories. Generically, other approaches fail to reproduce the universal early-time (free-streaming) attractor for the normalized longitudinal pressure $P_L/P$ in Bjorken flow, and the universal longitudinal and transverse free-streaming attractors for $\hat{P}_L$ and $\hat{P}_T$ at large negative and positive de Sitter times, respectively, in Gubser flow. Whenever the bulk and/or shear viscous stresses become large, the standard dissipative hydrodynamic approaches break down. The only other approach that can compete with ME hydrodynamics in the cases of Bjorken and Gubser flows is {\it anisotropic hydrodynamics}, but only because it uses for truncation of the Boltzmann moment hierarchy a custom-made ansatz of (modified) Romatschke-Strickland form for the distribution function that includes the momentum anisotropies associated with the shear and bulk viscous stresses in these flows non-perturbatively (albeit not by maximizing the Shannon entropy). Contrary to ME hydrodynamics, the success of anisotropic hydrodynamics for Bjorken and Gubser flows is therefore not expected to carry over to generic three-dimensional flow patterns.
 
We are confident that the excellent performance of ME hydrodynamics carries over to generic three-dimensional hydrodynamic evolution. To subject this confidence to rigorous numerical tests will require development of high-precision numerical solutions for (3+1)-dimensional kinetic theory for comparison. If successful, (3+1)-dimensional ME hydro will become the preferred macroscopic dynamical framework for relativistic heavy-ion collisions --- unless it turns out that the early stage of the latter is not sufficiently weakly coupled to admit some sort of kinetic description. Since ME hydro is based on a truncation of the Boltzmann moment hierarchy we do not know how to generalize it to fluids which are so strongly coupled that a kinetic theory approach becomes fundamentally inapplicable.

\section*{Acknowledgements}

We thank Jean-Paul Blaizot for stimulating discussions suggesting the exploration of entropy production in the ME hydro approach (see Appendix~\ref{appd}). CC thanks Sourendu Gupta for several illuminating discussions and insightful comments, and the organizers of the {\sl ETHCVM 2023} meeting (``Emergent Topics in Relativistic Hydrodynamics, Chirality, Vorticity, and Magnetic Fields'') in Puri, India, and of the {\sl Hot Quarks 2022} conference in Estes Park, Colorado, for providing an opportunity to present and discuss with participants parts of this work. Fruitful discussions with Derek Everett, Kevin Ingles, Lipei Du, Dananjaya Liyanage, Sunil Jaiswal, Amaresh Jaiswal, and Subrata Pal are also gratefully acknowledged. This work was supported by the U.S. Department of Energy, Office of Science, Office for Nuclear Physics under Awards No. \rm{DE-FG02-03ER41260} (C.C. and T.S.) and \rm{DE-SC0004286} (U.H.). Furthermore, the authors acknowledge partial support by the National Science Foundation under Grant No. NSF \rm{PHY-1748958} (KITP).


\appendix

\section{Sensitivity of kinetic theory solutions to the form of the initial distribution}
\label{appa}

\begin{figure}[h]
\includegraphics[width=0.8\linewidth]{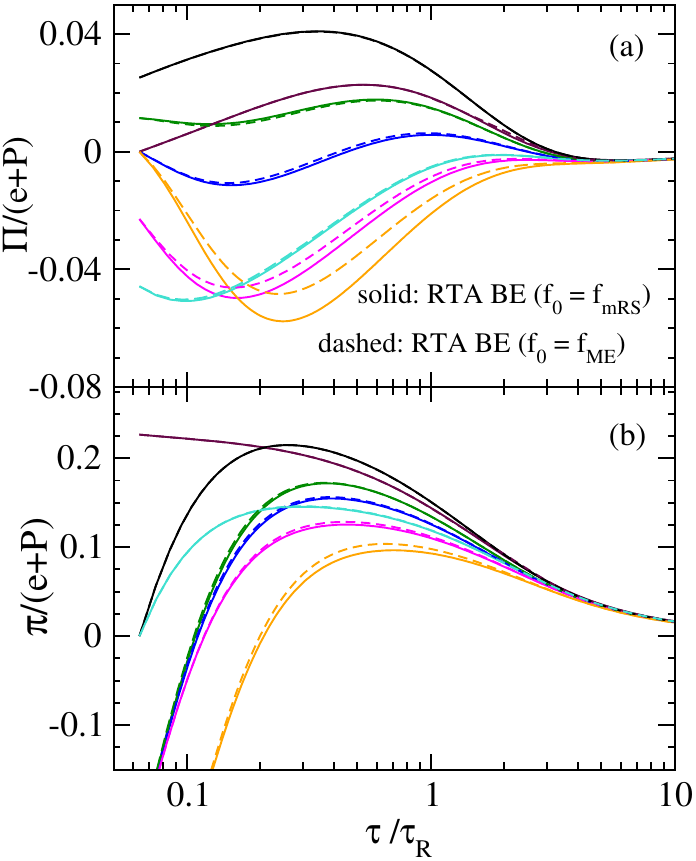}
\caption{
    Comparison of scaled bulk and shear stress evolution obtained from kinetic theory using an initial distribution of modified Romatschke-Strickland form (solid lines) and maximum-entropy type (dashed lines).
    \vspace*{-3mm}}\label{fig:appa1}
\end{figure}

In this appendix we test the sensitivity of non-conformal kinetic theory results to the choice of the initial distribution function used in the RTA Boltzmann equation. The solutions for a certain set of initial shear and bulk inverse Reynolds numbers (and temperature $T_0 = 500$ MeV) generated by $f_0 = f_\mathrm{mRS}$ have already been presented in Sec. \ref{non-conformal dynamics}. We now generate identical initial conditions for $(T_0, \bar\pi_0, \bar\Pi_0)$ using an initial distribution to be of maximum-entropy form, i.e., $f_0 = \fME$. The solutions are shown by dashed lines in Fig. \ref{fig:appa1} where the solid lines calculated with $f_0 = f_\mathrm{mRS}$ are the same as those in Fig. \ref{fig:pibar_Pibar_evol_NC}. The orange and magenta curves in Fig. \ref{fig:appa1}a show that non-hydrodynamic moments of the distribution function can lead to observable differences in the early-time evolution of hydrodynamic quantities. These differences are more prominent in the bulk channel than in the shear sector, and also for initial conditions where the shear inverse Reynolds numbers are large and negative ($\bar\pi_0 = - 0.41$ for the orange curve and $- 0.22$ for the magenta one). For $\bar\pi_0 \geq 0$, the differences are found to be negligible.

\begin{figure}[t]
\includegraphics[width=0.8\linewidth]{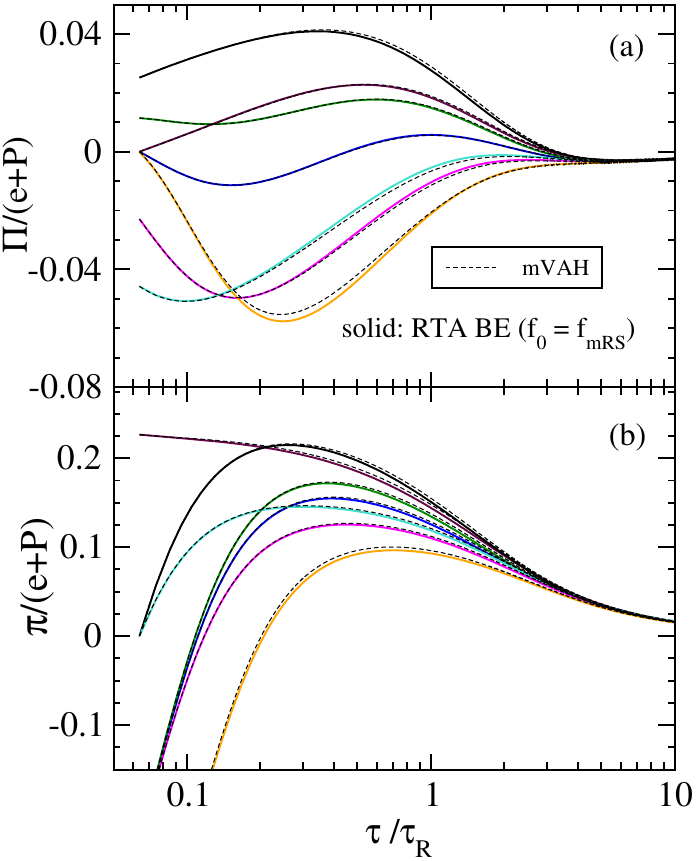}
\caption{
    Comparison of dissipative flux evolution obtained from kinetic theory using an initial distribution $f_0 = f_\mathrm{mRS}$ (solid lines) and modified viscous anisotropic hydro (mVAH) (black dashed lines).
    }\label{fig:appa2}
\end{figure}

\begin{figure}[htb!]
\includegraphics[width=0.8\linewidth]{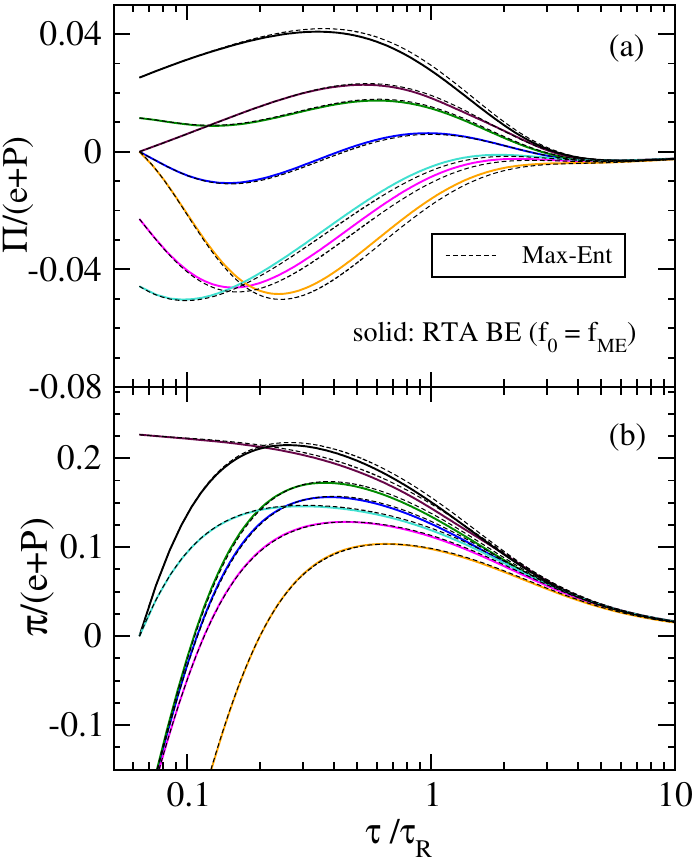}
\caption{
    Comparison of scaled bulk and shear evolution obtained from kinetic theory using $f_0 = \fME$ (solid lines) and ME-hydro (black dashed lines). 
        }\label{fig:appa3}
    \end{figure}
 
Nevertheless, the early time non-universality in kinetic theory results corresponding to the two choices of initial distribution suggest that for $f_0 = f_\mathrm{mRS}$ modified anisotropic hydro might perform better than ME-hydro, and vice-versa for $f_0 = \fME$. We verify this intuition in Figs.~\ref{fig:appa2} and \ref{fig:appa3}. In Fig.~\ref{fig:appa2} the comparison between kinetic theory solutions for $f_0 = f_\mathrm{mRS}$ (shown by solid lines) with m-aHydro (black dashed lines) shows that the two are in excellent agreement. A similar comparison in Fig.~\ref{fig:appa3}  between kinetic theory solutions with $f_0{\,=\,}\fME$ (shown by {\it solid} lines\footnote{%
    Note that the solid lines of Fig.~\ref{fig:appa3} are identical to the dashed lines of the corresponding color in Fig.~\ref{fig:appa1}. 
}) and ME-hydro (black dashed lines) shows that the latter describes the former well. Thus, for some initial conditions the band of uncertainty in the microscopic description of hydrodynamic quantities due to an incomplete knowledge of the initial distribution may be too large to distinguish between the superiority of one macroscopic description over another.

\section{Classical vs quantum statistics}
\label{appb} 

\begin{figure}[t]
\includegraphics[width=0.8\linewidth]{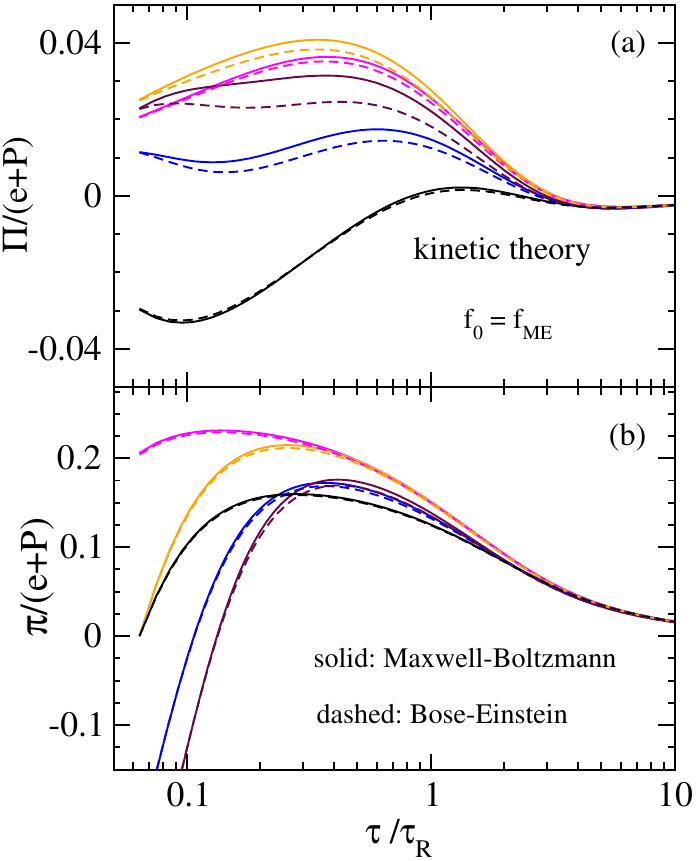}
\caption{
    Comparison of scaled bulk and shear evolution obtained from kinetic theory for particles obeying Bose-Einstein statistics (dashed) and classical statistics (solid).
    \vspace*{-3mm}}\label{fig:appb1}
\end{figure}

\begin{figure}[t]
\includegraphics[width=0.8\linewidth]{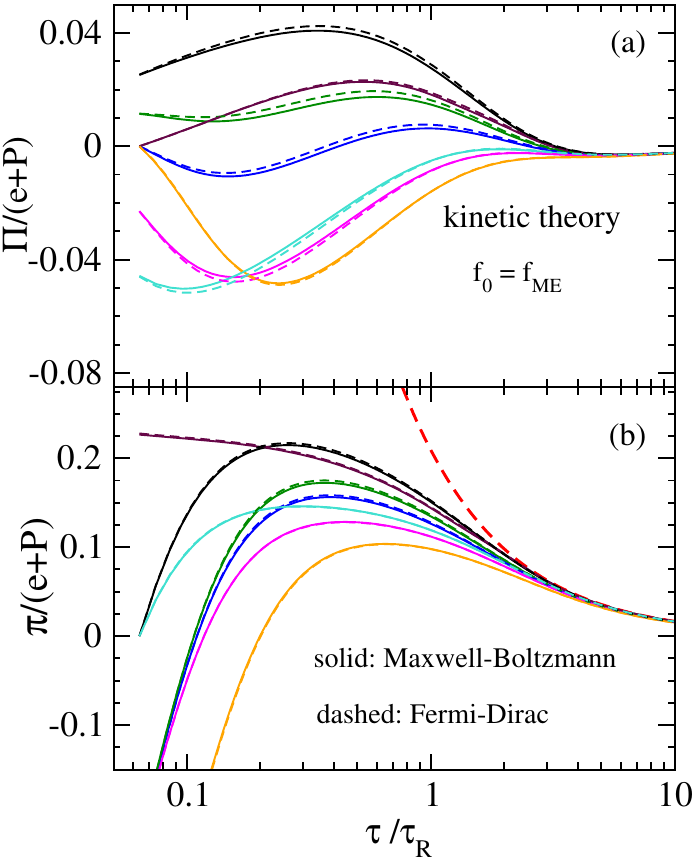}
\caption{
    Same as Fig.~\ref{fig:appb1}, but for dashed lines now standing for Fermi-Dirac statistics.
    \vspace*{-3mm}}\label{fig:appb2}
\end{figure}

In Figs.~\ref{fig:appb1} and \ref{fig:appb2} we explore the extent to which the Maxwell-Boltzmann approximation describes the macroscopic dynamics of particles governed by quantum statistics when subject to Bjorken flow.\footnote{%
    For the exact solution of the (0+1)-dimensional RTA Boltzmann equation for a gas of massive Bose-Einstein or Fermi-Dirac particles see \cite{Florkowski:2014sda}.} For both of these figures we use initial distributions of maximum-entropy type, i.e., $f_0$ given by Eq.~(\ref{fME_LRF}) for classical statistics and by Eq.~(\ref{MaxEnt distribution quantum}) for quantum statistics with appropriate choice of $\theta$. The results for classical and quantum statistics are denoted, respectively, by solid and dashed lines. The blue, maroon, magenta, and orange curves in Fig. \ref{fig:appb1}a shows that Boltzmann approximation does not appropriately capture the early-time bulk viscous pressure evolution for particles obeying Bose-Einstein statistics. In contrast, for Fermi-Dirac particles the agreement between classical and quantum statistics is much better, as demonstrated in Fig. \ref{fig:appb2}.  

\begin{figure}[h]
\includegraphics[width=0.8\linewidth]{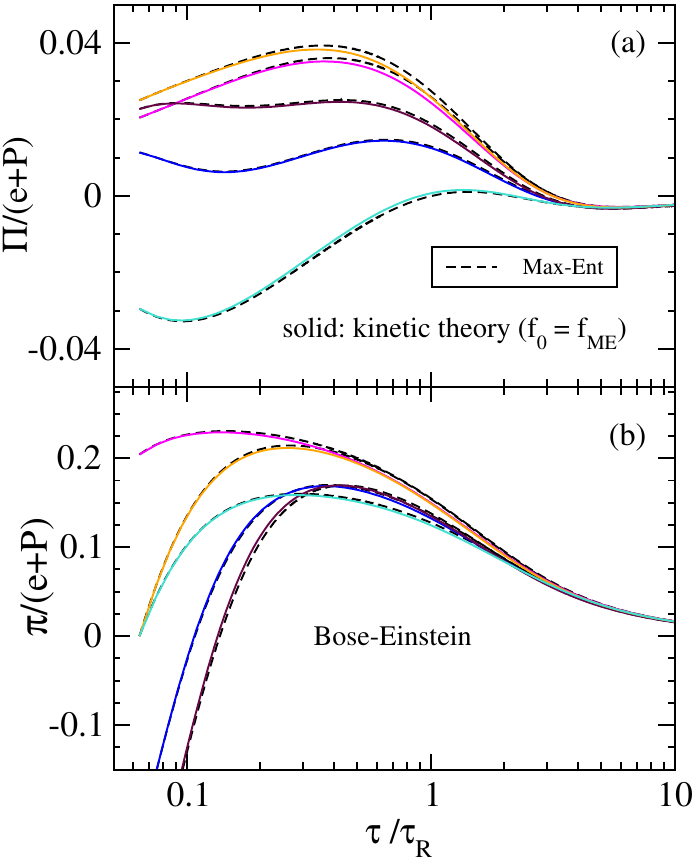}
\caption{
    Comparison between kinetic theory and ME-hydro for Bose-Einstein statistics.
    \vspace*{-3mm}}
    \label{fig:appb3}
\end{figure}

\begin{figure}[h]
\includegraphics[width=0.8\linewidth]{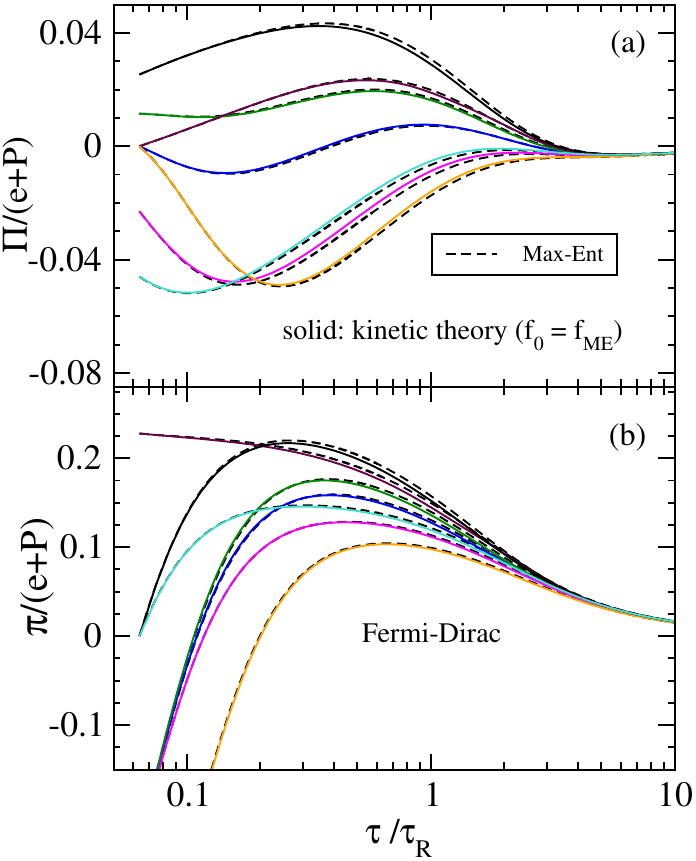}
\caption{
    Same as Fig. \ref{fig:appb3} but for Fermi-Dirac statistics.
    \vspace*{-3mm}}
    \label{fig:appb4}
\end{figure}

In Figs.~\ref{fig:appb3} and \ref{fig:appb4} we show ME-hydro solutions for Bose-Einstein and Fermi-Dirac statistics, respectively, and compare with the corresponding exact kinetic theory solutions. For quantum ME hydro one must solve Eqs.~(\ref{e_evol_ME}-\ref{PT_evol_ME}) using $f = \fME$ given by Eq.~(\ref{MaxEnt distribution quantum}) to obtain the moments $\tilde{I}_{nrq}$ (see Eq.~(\ref{I_ME})). As for the Boltzmann case, we solved directly for the Lagrange parameters $(\Lambda, \lambda_\Pi, \gamma)$. The Jacobian of transformation relating $M^{a}_{\,\,b}$ to the moments $\tilde{I}_{nrq}$ (Eqs. \ref{M_row1}-\ref{M_row3}) is the same as for classical statistics, up to changing $\tilde{I}_{nrq} \to \tilde{I}^\mathrm{q}_{nrq}$ where
\begin{align}
\tilde{I}^\mathrm{q}_{nrq} = \frac{1}{(2q)!!} \int dP \, \left(p^\tau\right)^{n-r-2q} \, \left(\frac{p_\eta}{\tau} \right)^r \, p^{2q}_{T} \, \fME \, \tilde{f}_{\mathrm{ME}},   
\end{align}
with $\tilde{f}_{\mathrm{ME}} \equiv 1{\,-\,}\theta\, \fME$. Figs.~\ref{fig:appb3} and \ref{fig:appb4} demonstrate that ME-hydro gives an excellent description of non-conformal kinetic theory for quantum statistics as well.  

\section{Evolution of the Lagrange parameters}
\label{appc} 

\begin{figure}[h]
\begin{center}
\includegraphics[width=0.85\linewidth]{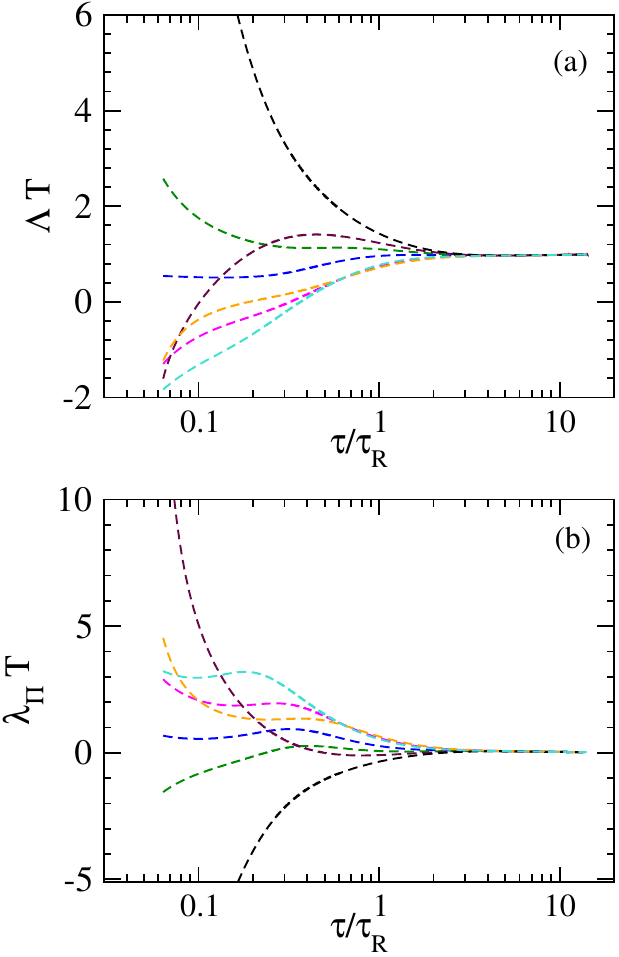}
\end{center}
\vspace{-0.3cm}
\caption{Evolution of Lagrange parameters $\Lambda$ and $\lambda_\Pi$ (both in units of the instantaneous inverse temperature) of the maximum-entropy distribution function corresponding to different initial conditions; see text for details. 
    }
    \label{fig:appc1}
\end{figure}

\begin{figure}
\begin{center}
\includegraphics[width=0.85\linewidth]{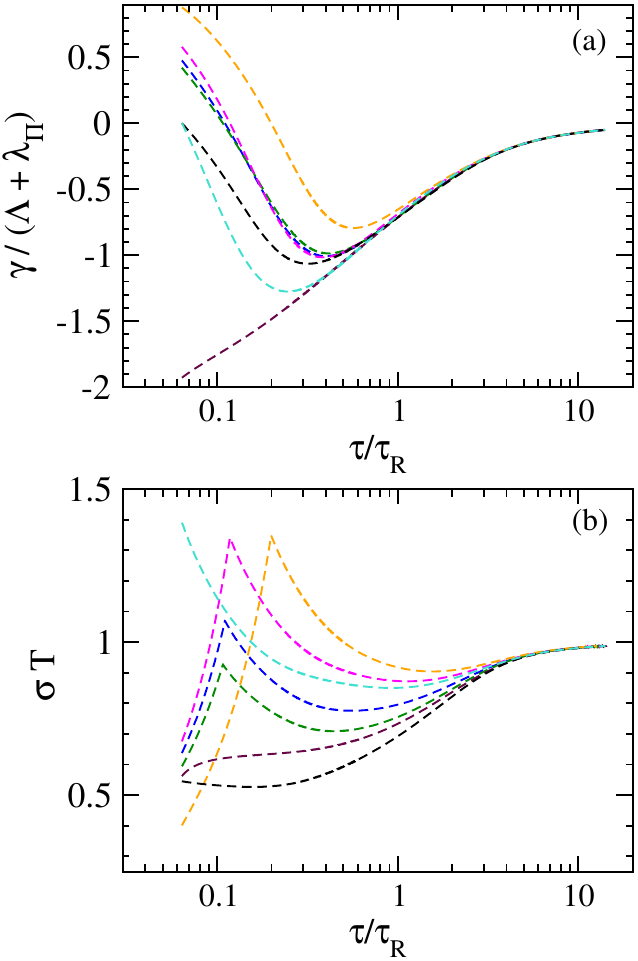}
\end{center}
\vspace{-0.3 cm} 
\caption{Evolution of the scaled Lagrange parameter for shear stress, $\gamma/(\Lambda + \lambda_\Pi)$, in panel (a). Panel (b) shows the evolution of parameter $\sigma \equiv \Lambda + \lambda_\Pi - |\mathrm{min}(\gamma/2,-\gamma)|$ for different initial conditions; see text for details.
    }
    \label{fig:appc2}
\end{figure}

We discuss the evolution of the ME Lagrange parameters in ME hydrodynamics for non-conformal fluids undergoing Bjorken flow. Similar to Fig.~\ref{fig:LP_conformal} showing the corresponding conformal evolution, we explore in Figs.~\ref{fig:appc1} and \ref{fig:appc2} the time dependence of the Lagrange parameters $(\Lambda, \lambda_\Pi, \gamma)$ (suitably normalized as mentioned later) controlling the ME distribution function in a non-conformal gas. In Figs.~\ref{fig:appc1}a we notice a curious feature that the parameter $\Lambda$, which one may be inclined to associate with an inverse temperature, is negative! This is not a problem though, as for the maximum-entropy distribution to be well-behaved $\Lambda$ does not need to be positive. Instead, it is the sum of $\Lambda$ and $\lambda_\pi$ that must be positive (see Eq.~(\ref{bound_Bjorken})). In fact, negative values of $\Lambda$ arise as our initial conditions correspond to negative bulk viscous pressures for a gas at moderate temperatures ($T_0/m = 1$). 

One observes from Fig.~\ref{fig:appc1}a that the magenta, maroon, orange, and cyan curves that correspond to initial $\Pi/P \leq 0$ and substantial initial shear stress are characterised by negative $\Lambda_0$. In contrast, the blue, green, and black curves with initial $\Pi/P \geq 0$ have positive $\Lambda_0$. These are in line with the arguments given above. The various solutions in panel (a) evolve distinctly from each other during the early stages. At times $\tau \approx 2 \, \tau_R$ they approach unity suggesting onset of near-equilibrium dynamics when $\Lambda$ can be thought of as an inverse temperature. The solutions for $\lambda_\Pi T$, too, evolve quite differently from each other at early times before approaching zero as the bulk viscous pressure of the system vanishes. Note that for all curves, the sum $\Lambda + \lambda_\Pi$ is greater than zero; see also Table \ref{table_ic_Lagrange}.

In Fig.~\ref{fig:appc2}a we plot the evolution of Lagrange multiplier $\gamma$ (made dimensionless by dividing by $(\Lambda{\,+\,}\lambda_\Pi)$) which controls the momentum space anisotropy of the distribution. As the black and cyan curves start with vanishing shear stress, they correspond to $\gamma_0{\,=\,}0$. The maroon curve has a large initial positive shear (or $P_L/P \ll 1$) and starts with a large negative initial $\gamma$. Similar to the $\Lambda T$ and $\lambda_\Pi T$ evolution in Fig.~\ref{fig:appc1}, the far-off-equilibrium dynamics of $\gamma/(\Lambda + \lambda_\Pi)$ are strongly dependent on initial conditions. However, at late times, all of them approach zero as the system isotropizes. We finally plot in Fig.~\ref{fig:appc2}b the time evolution of $\sigma \equiv \Lambda + \lambda_\Pi - \big|\mathrm{min}(\gamma/2, -\gamma)\big|$ (again, made dimensionless by multiplication with the temperature) for all the ME hydro solutions discussed so far. According to Eq.~(\ref{bound_Bjorken}) $\sigma$ must be positive throughout the system's evolution for the Maximum Entropy distribution to be well-behaved. We see that this is indeed the case for all curves. The kink in the magenta, green, orange and blue curves occurs when $\gamma$ crosses zero. At late times, $\lambda_\Pi$ and $\gamma$ approach zero, and $\sigma$ assumes the role of an inverse temperature such that $\sigma T \approx 1$.   

\section{Entropy evolution}
\label{appd} 

\begin{figure}
\begin{center}
\includegraphics[width=0.95\linewidth]{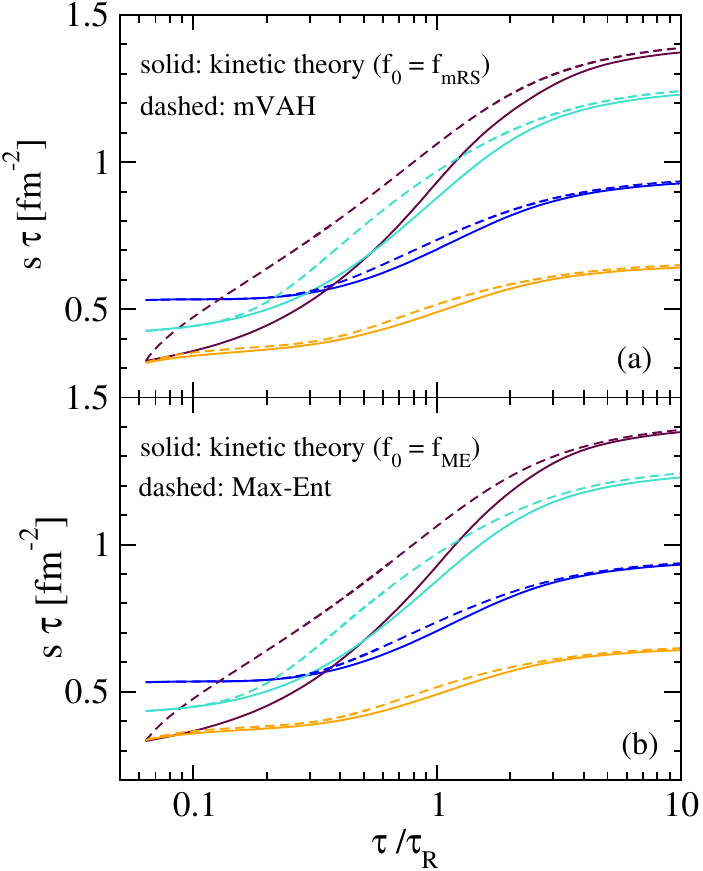}
\end{center}
\vspace{-0.3 cm} 
\caption{Evolution of entropy per unit rapidity and transverse area in non-conformal Bjorken flow. Solid lines are exact solutions of kinetic theory; dashed lines correspond to hydrodynamic approximations. For the color coding, refer to Table \ref{table:IC}. 
    }\label{fig:appd1}
\end{figure}

It has been shown earlier that both $\mVAH$ and ME-hydro describe the kinetic evolution of \textit{hydrodynamic} moments of the distribution function rather well. In this appendix we compare their performance when it comes to modeling a non-hydrodynamic moment of $f$. In Fig.~\ref{fig:appd1} we plot the evolution of the non-equilibrium entropy per unit rapidity and transverse area, $s \tau$, obtained using kinetic theory and the hydrodynamic approximations.\footnote{%
    For non-dissipative Bjorken expansion, the entropy of a fluid element does not change with time which manifests in $s \tau$ being a constant of motion.}

In panel (b) we solve kinetic theory with an initial distribution $f_0 = \fME$ that generates $(\tilde{\Pi}_0, \tilde{\pi}_0)$ as mentioned in Table~\ref{table:IC}, and then compute the entropy density from Eq.~(\ref{H-function}). The ME-hydro results for these initial conditions were already obtained in Appendix~\ref{appa}, Fig.~\ref{fig:appa3}. We use the corresponding solutions for $(\Lambda, \lambda_\Pi, \gamma)$ to calculate the ME-hydro entropy density (\ref{entropy_Boltzmann}). The corresponding results for $s \tau$ are shown by dashed lines in Fig.~\ref{fig:appd1}b. For panel (a) we repeat the same procedure but use $f_0 = f_\mathrm{mRS}$ for the kinetic theory solutions, and for $\mVAH$ we calculate the entropy density (\ref{H-function}) by plugging in the modified Romatschke-Strickland distribution (\ref{f_in}) with the corresponding time-dependent RS parameters \cite{Jaiswal:2021uvv}. 

Both panels show substantial differences in evolution between the exact non-equilibrium entropy and the hydrodynamic results. From the analysis of Calzetta and Cantarutti \cite{Calzetta:2019dfr} it can be deduced that in the conformal case the divergence of the entropy four-current for a maximum-entropy type distribution is given by,
\begin{align}
    \partial_\mu S^{\mu} = - \frac{1}{\tau_R} \, \gamma_{\langle \mu\nu \rangle} \, \pi^{\mu\nu}.
\end{align}
In the non-conformal case $\partial_\mu S^\mu$ gets generalized to,
\begin{align}
    \partial_\mu S^{\mu} = - \frac{1}{\tau_R} \, \left( \gamma_{\langle \mu\nu \rangle} \, \pi^{\mu\nu} + 3 \, \lambda_\Pi \, \Pi  \right).
\end{align}
For a system undergoing Bjorken expansion this implies that the entropy per unit rapidity increases as
\begin{align}
    \frac{d\left(s\tau\right)}{d\tau} = - 3 \bar\tau \, \left( \gamma \, \frac{\pi}{2} + \lambda_\Pi \, \Pi \right), \label{slope_entropy_MEhydro}
\end{align}
where $\bar\tau \equiv \tau/\tau_R$. This can also be verified using the entropy density definition (\ref{entropy_Boltzmann}):
\begin{align}
    s = \Lambda e + \lambda_\Pi \, \left( P_L + 2 P_T \right) + \gamma \, \left( P_T - P_L \right) + n. 
\end{align}
Taking the differential of $s$ and noting that
\begin{align}
    d n = - e \, d\Lambda - \left( P_L + 2 P_T \right) \, d \lambda_\Pi - \left( P_T - P_L \right) \, d\gamma
\end{align}
one finds\footnote{%
    Note that for a system in equilibrium $ds = \beta \, de$ as expected.
}
\begin{align}
    ds = \Lambda \, de + \lambda_\Pi \, \left(dP_L + 2 \, dP_T\right) + \gamma \, \left(dP_T - dP_L\right).
\end{align}
Now, using the ME-hydro equations of motion (\ref{e_evol_ME}-\ref{PT_evol_ME}) we obtain
\begin{align}
    \frac{d\left(s\tau\right)}{d\tau} &= - 3 \bar\tau \left( \lambda_\Pi \, \Pi + \gamma \, \frac{\pi}{2} \right) + n - \Lambda \, P_L  \, \nonumber\\
    & - \gamma \, \left(\tilde{I}_{240} - 2  P_L - \tilde{I}_{221} \right) \nonumber\\
    & - \lambda_\Pi \left( 2 P_L - \tilde{I}_{240} - 2 \tilde{I}_{221} \right). \label{stau_ME_hydro1}
\end{align}
In order to show that the terms in r.h.s. of above equation that are not proportional to $\bar\tau$ cancel each other we start with the number density (in local rest frame coordinates):
\begin{align}
    n &= \int_0^{\infty} \frac{dp_T \, p_T}{2\pi^2} \, \int_0^{\infty} dp_z \, \fME \nonumber\\
    &= - \int_0^{\infty} \frac{dp_T \, p_T}{2\pi^2} \, \int_0^{\infty} dp_z \, p_z \, \frac{\partial \fME}{\partial p_z}. \label{n_inte_by_parts}
\end{align}
We then use,
\begin{align}
    p_z \, \frac{\partial \fME}{\partial p_z} &= - \fME \, \frac{p_z^2}{E_p} \, \Bigg[ \Lambda + \lambda_\Pi \, \left( 2 - \frac{p_T^2 + p_z^2}{E_p^2} \right) \Bigg] \nonumber \\
    & + \gamma \, \left( - 2 - \frac{p_T^2}{2 \, E_p^2} + \frac{p_z^2}{E_p^2} \right) \Bigg],
\end{align}
to obtain,
\begin{align}
    n & = \Lambda \, P_L + \lambda_\Pi \left( 2 P_L - \tilde{I}_{240} - 2 \tilde{I}_{221} \right) \nonumber \\
     & + \gamma \, \left( \tilde{I}_{240} - 2 P_L - \tilde{I}_{221} \right).
\end{align}
Accordingly, the entropy production rate (\ref{stau_ME_hydro1}) stems solely from terms proportional to $\bar\tau$ (or the collisional kernel):
\begin{align}\label{stau_first_derivative_ME}
    \frac{d\left(s\tau\right)}{d\tau} = - 3 \bar\tau \left( \lambda_\Pi \, \Pi + \gamma \, \frac{\pi}{2} \right),
\end{align}
consistent with Eq.~(\ref{slope_entropy_MEhydro}). Eq.~(\ref{slope_entropy_MEhydro}) also implies that the initial slopes of the solid and dashed curves in Fig.~\ref{fig:appd1}b must be equal. This is because the kinetic theory curves are initialised with a maximum-entropy type distribution such that
\begin{align}
    \left.\frac{d\left(s\tau\right)}{d\tau}\right\vert_{\tau_0} &= \left(\frac{\tau}{\tau_R}\right)_0 \int dP \, p^\tau \, \log(\fME) \, \left(\fME - f_{eq,0}\right), \nonumber\\
    &= - 3 \bar\tau_0 \left( \lambda_{\Pi,0} \, \Pi_0 + \gamma_0 \, \frac{\pi_0}{2} \right).
\end{align}
Although in Fig.~\ref{fig:appd1}b the solid and dashed curves in orange, blue, and cyan colors graze along each other at early times, the initial slopes of the solid and dashed maroon curves differ strongly. In fact, the initial slope of the maroon ME-hydro curve agrees with the expectation (\ref{slope_entropy_MEhydro}) whereas the kinetic theory solution does not; the rate of entropy production is considerably smaller for the latter than for the former. To understand this puzzling behavior we compute the second time derivative of $s\tau$ in kinetic theory. Starting from (\ref{H-function}) multiplied by $\tau$ and evaluating the first time derivative using the RTA Boltzmann equation gives
 \begin{align}\label{entropy_pr_KT}
    \frac{d(s\tau)}{d\tau} = \frac{\tau}{\tau_R} \, \int dP \, p^\tau \, \log(f) \, \delta f
 \end{align}
 where $dP = d^2p_T dp_\eta/[(2\pi)^3 \tau p^\tau]$. One then obtains for the second derivative
\begin{align}
    \frac{d^2(s\tau)}{d\tau^2} &= - \frac{d(s\tau)}{d\tau} \, \left( \frac{1}{\tau_R} + \frac{d\log \tau_R}{d\tau} \right) 
\nonumber \\
    & - \frac{\tau}{\tau_R} \, \int dP \, p^\tau \, \log(f) \, \frac{\partial\feq}{\partial \tau} 
\nonumber \\
    & - \frac{\tau}{\tau_R^2} \, \int dP \, p^\tau \, \frac{(\delta f)^2}{f}. \label{stau_second_derivative}
\end{align}
While the first momentum integral appearing on the r.h.s. involving the derivative of $\feq$ is well behaved, it is not obvious that the last one converges, especially for large deviations from equilibrium. On expanding the integrand, $(\delta f)^2/f = f - 2 \feq + \feq^2/f$, it becomes clear that for the initial slope of the entropy production rate to be finite, the quantity $f_{\mathrm{eq},0}^2/\fME$ must decay to zero at large momenta. This in turn requires
\begin{align}
    \frac{2}{T_0} - \Lambda_0 - \lambda_{\Pi,0} > \left|\mathrm{min} \left( - \frac{\gamma_0}{2}, \gamma_0 \right)\right|.
\end{align}
This criterium is not met by the maroon curve. In consequence the initial slope of the kinetic entropy production rate has a singularity: the slope of $d(s\tau)/d\tau$ approaches negative infinity. This singularity is not captured by ME hydro. To pin-point the origin of this difference let us calculate the quantity $d^2(s\tau)/d\tau^2$ in ME hydro. Re-writing Eq.~(\ref{stau_first_derivative_ME}) as 
\begin{align}
    \frac{d(s\tau)}{d\tau} =  \frac{\tau}{\tau_R} \, \int dP \, p^\tau \, \log(\fME) \, \left( \fME - \feq \right)
\end{align}
and taking another time derivative we have
\begin{align}\label{stau_second_derivative_ME1}
    \frac{d^2(s\tau)}{d\tau^2} &= - \frac{d\log \tau_R}{d\tau} \, \frac{d(s\tau)}{d\tau} \nonumber \\
    & + \frac{\tau}{\tau_R} \, \int dP \, p^\tau \, \log(\fME) \, \frac{\partial \fME}{\partial \tau} \nonumber \\
    & - \frac{\tau}{\tau_R} \, \int dP \, p^\tau \, \log(\fME) \, \frac{\partial \feq}{\partial \tau} \nonumber \\
    & + \frac{\tau}{\tau_R} \, \int dP \, p^\tau \, \frac{\left(\fME - \feq \right)}{\fME} \, \frac{\partial \fME}{\partial \tau}.
\end{align}
We can simplify the second term on the r.h.s. of this equation as follows: although $\fME$ is not a solution of the Boltzmann equation, the ME hydro equations of motion (\ref{e_evol_ME}-\ref{PT_evol_ME}) permit replacement of the partial derivative $\partial \fME/\partial \tau$ by $-(\fME - \feq)/\tau_R$. To see this we write 
\begin{align}\label{d2stau_dtau2_term2}
    & \int dP \, p^\tau \log(\fME) \, \frac{\partial \fME}{\partial \tau} = - \int dP \, \frac{\partial \fME}{\partial \tau} 
\nonumber \\
    & \times \Bigl[ \Lambda \, (p^\tau)^2 + \lambda_\Pi \, \Bigl( p_T^2 + \frac{p_\eta^2}{\tau^2}  \Bigr) + \gamma \, \Bigl(\frac{p_T^2}{2} - \frac{p_\eta^2}{\tau^2} \Bigr) \Bigr]
 \end{align}
and use the ME hydro equations of motion for $(e, P_L, P_T)$ in the form \cite{Calzetta:2019dfr}
\begin{align}
    \int dP \, (p^\tau)^2 \, \Bigl[ \frac{\partial \fME}{\partial \tau} + \frac{1}{\tau_R} \, \left(\fME - \feq \right) \Bigr] &= 0 ,
\nonumber \\
     \int dP \, \frac{p_\eta^2}{\tau^2} \, \Bigl[ \frac{\partial \fME}{\partial \tau} + \frac{1}{\tau_R} \, \left(\fME - \feq \right) \Bigr] &= 0,
\nonumber \\
      \int dP \, \frac{p_T^2}{2} \Bigl[ \frac{\partial \fME}{\partial \tau} + \frac{1}{\tau_R} \, \left(\fME - \feq \right) \Bigr] & = 0.
\end{align}
The second term on the r.h.s. of (\ref{stau_second_derivative_ME1}) can therefore be written as\footnote{%
    Note that the term involving the Lagrange multiplier $\Lambda$ in Eq. (\ref{d2stau_dtau2_term2}) vanishes due to Landau matching.
    }
\begin{align}
      - \frac{\tau}{\tau_R^2} \, \int dP \, p^\tau  \log(\fME) \, \left(\fME - \feq \right) = - \frac{1}{\tau_R} \, \frac{d(s\tau)}{d\tau},
\end{align}
and the second-derivative of $s\tau$ in ME hydro becomes
\begin{align}\label{stau_second_derivative_ME2}
    \frac{d^2(s\tau)}{d\tau^2} &=  - \frac{d(s\tau)}{d\tau} \, \left( \frac{1}{\tau_R} + \frac{d\log \tau_R}{d\tau} \right) 
\nonumber \\
    & - \frac{\tau}{\tau_R} \, \int dP \, p^\tau \, \log(\fME) \, \frac{\partial\feq}{\partial \tau} 
\nonumber \\
    & + \frac{\tau}{\tau_R} \, \int dP \, p^\tau \, \frac{\left(\fME - \feq \right)}{\fME} \, \frac{\partial \fME}{\partial \tau}. 
\end{align}
At $\tau = \tau_0$ and for kinetic theory initialised with a maximum entropy distribution, the r.h.s. of the above expression agrees with the kinetic result (\ref{stau_second_derivative}) for the first two terms.\footnote{%
    Note that the initial slope of the temperature in ME hydro is identical to that in kinetic theory initialised with $\fME$. This also implies that $\partial \feq/\partial \tau$ at $\tau = \tau_0$ is identical in both approaches.
} 
However, as $\fME$ does not solve the Boltzmann equation, the last term differs in the two approaches:
\begin{align}
    \frac{\left(\fME - \feq \right)}{\fME} \, \frac{\partial \fME}{\partial \tau} \neq - \frac{1}{\tau_R} \, \frac{\left(\fME - \feq\right)^2}{\fME}.
\end{align}
Note that the source of divergence for $d^2(s\tau)/d\tau^2$ at $\tau{\,=\,} \tau_0$ is precisely the term $(\fME - f_{eq,0})^2/\fME$. In ME hydro $\partial \fME/\partial \tau \propto \fME$ such that
\begin{align}
    \frac{\left( \fME - \feq \right)}{\fME} \, \frac{\partial \fME}{\partial \tau} \propto  \left( \fME - \feq \right)
\end{align}
whose momentum integral in (\ref{stau_second_derivative_ME2}) is finite. This is why ME hydro does not capture the singularity in the second time derivative of $s\tau$ for initial conditions deviating strongly from thermal equilibrium.\footnote{%
    We found that the discrepancies between ME hydro and kinetic theory typically become serious once $P_L/P$ or $P_T/P$ drop below about ten percent.
}

\section{Gubser flow in Milne coordinates}
\label{appe} 

Although it is mathematically convenient to work with Gubser flow in curved space-time parametrized by de-Sitter coordinates, one may gain a more physical picture of the expanding fluid by expressing it in familiar Milne coordinates. For instance, the quantity $\hat{\theta} \equiv 2 \tanh\rho$ which is usually dubbed as the scalar expansion rate in Gubser flow goes negative at $\rho < 0$, although the fluid is always expanding in flat space. It should be noted that unlike scalar quantities such as temperature where the transformation from Milne to Gubser coordinates is implemented by a simple scaling, $T \to \hat{T} = \tau T$, the relation between the Milne expansion rate $\theta$ and the Gubser expansion rate $\hat{\theta}$ is more complicated (see Eq.~(7) of \cite{Loganayagam:2008is}):
\begin{align}\label{theta_Gubser_Milne}
    \tau \theta = \hat{\theta} + \frac{3}{\tau} \, \frac{\partial \tau}{\partial \rho}  
     = \frac{1 + 2 \tilde{r}^2 + 5 \tilde{\tau}^2}{s_{+} \, s_{-}}.
\end{align}
Here $s_{\pm} \equiv (1+r_{\pm}^2)^{1/2}$, with $r_{\pm} \equiv (\tilde{r} \pm \tilde{\tau})$, and the scaled coordinates are $\tilde{r}\equiv q\,r$ and $\tilde{\tau} = q\, \tau$ (here, $q$ characterizes the inverse size of the system). To get to the last equality in (\ref{theta_Gubser_Milne}) we used the relations (see Eq.~(148) of \cite{Gubser:2010ui})
 \begin{align}
     \tilde{\tau} = \frac{\mathrm{sech}\rho}{\cos\theta - \tanh\rho}, \qquad 
     \tilde{r} = \frac{\sin\theta}{\cos\theta - \tanh\rho},
 \end{align}
and wrote everything in terms of Milne coordinates.\footnote{%
    The $\theta$ appearing in Gubser coordinates $(\rho, \theta, \phi,\eta)$ should not be confused with the scalar expansion rate, also denoted by the same symbol.
}
Note that Eq.~(\ref{theta_Gubser_Milne}) could have also been obtained directly from the definition of the expansion rate in flat space-time:
\begin{align}
    \theta \equiv d_\mu u^\mu = \frac{\partial u^\tau}{\partial \tau} + \frac{\partial u^r}{\partial r} + \frac{u^\tau}{\tau} + \frac{u^r}{r},
\end{align}
where, in Milne coordinates, the components of $u^\mu$ are given by \cite{Gubser:2010ui} $u^\tau \equiv \cosh\kappa$, $u^r \equiv \sinh\kappa$, with
\begin{align}
    \tanh\kappa(\tau, r) = \frac{2 \tilde{\tau} \tilde{r}}{1 + \tilde{\tau}^2 + \tilde{r}^2}. 
\end{align}
In order to extract the early and late-time behaviors of the flow, we compute the longitudinal and transverse expansion rates \cite{McNelis:2021zji}, $\theta_L \equiv z_\mu d_z u^\mu, \,\, \theta_\perp \equiv \theta - \theta_L$,
where $d_z = - z^\mu d_\mu$, with $z^\mu$ being the longitudinal component of velocity,
\begin{align}
    z^\mu = \frac{1}{\sqrt{1 + \left(u^r \right)^2}} \, \left(\tau u^\eta, 0, 0, \frac{u^\tau}{\tau} \right).
\end{align}
For Gubser flow $u^\eta = 0$ such that the only non-vanishing component of $z^\mu$ is $z^\eta = 1/\tau$. Accordingly the longitudinal expansion rate is
\begin{align}
    \theta_L = - z_\eta z^\eta \, \left( \frac{\partial u^\eta}{\partial \eta}  + \Gamma^{\eta}_{\eta\tau} \, u^\tau \right) = \frac{\cosh\kappa}{\tau}.
\end{align}
Written in terms of Milne variables, the longitudinal and transverse expansion rates are
\begin{align}\label{theta_L_theta_p}
    \tau \theta_L = \frac{1 + \tilde{r}^2 + \tilde{\tau}^2}{s_{+} s_{-}}, 
    \qquad 
    \tau \theta_\perp = \frac{\tilde{r}^2 + 4 \tilde{\tau}^2}{s_{+} s_{-}}.
\end{align}
The symmetries of Gubser flow imply that the state of a fluid element $A$ located at $r_A$ at proper time $\tau_A$ is identical to that of the central fluid cell $C(r{=}0)$ at a proper time $\tau_C = (1/q) \times \mathrm{sech}\rho_A/(1{-}\tanh\rho_A)$, where $\rho_A = \rho(\tau_A, r_A)$. This allows us to focus on the central cell. Eq.~(\ref{theta_L_theta_p}) implies that the longitudinal expansion rate of Gubser flow is $\theta_L = 1/\tau$ at all times, which is identical to the Bjorken expansion rate, as expected. In contrast, the transverse expansion rate $\theta_\perp = (1/\tau) \times 4 \tilde{\tau}^2/(1 + \tilde{\tau}^2)$ vanishes at early times $\tau \ll 1/q$ but approaches $\theta_\perp \approx 4/\tau$ at late times. Clearly, $\theta_\perp$ dominates over $\theta_L$ at late times, with a transition taking place around $\tilde{\tau} = \sqrt{1/3}$. 
Note that the corresponding Gubser transition time is $\rho \approx - 0.55$. 

Although the transverse expansion rate exceeds the longitudinal one at late times, it decreases with $\tau$. Why then does the system not thermalize at late times as in Bjorken flow? The reason is the following: once $P_T/e \ll 1$, the temperature decreases as, $d\hat{T}/d\rho = - (\tanh\rho)  \, \hat{T}/2$. For the central cell, $\partial_\rho = \tau \, \partial_\tau$, such that at late times (or large $\rho$),
\begin{align}
    \frac{\partial T}{\partial \tau} \approx -\frac{3}{2} \, \frac{T}{\tau} \implies T \sim \tau^{-3/2}.
\end{align}
Accordingly, the expansion rate $\theta \propto 1/\tau$ exceeds the microscopic scattering rate $1/\tau_R \propto T \sim \tau^{-3/2}$ and does not permit the system to thermalize. Note that assuming (incorrectly) an ideal cooling law for the central cell leads to even faster cooling at late times, $T \sim \tau^{-5/3}$ (hence, an even slower microscopic scattering rate), leading to the same conclusion.

\bibliography{references.bib}

\end{document}